\documentclass[aps,prd,preprintnumbers,showpacs,showkeys,nofootinbib,superscriptaddress,fleqn,floatfix,tightenlines,10pt]{revtex4-2} 
\usepackage{amsmath,amsfonts,amssymb,amscd,amsxtra,amsthm}
\usepackage{dsfont}
\usepackage{graphicx}  
\usepackage{epstopdf}
\usepackage{dcolumn}  
\usepackage{bm}          
\usepackage{slashed}
\usepackage[utf8]{inputenc}
\usepackage{booktabs}
\usepackage[normalem]{ulem} 
\usepackage[dvipsnames]{xcolor} 
\usepackage{enumitem}
\usepackage{array}
\usepackage{slashed}
\usepackage{tikz}
\usepackage{float}
\usepackage{multirow}
\usepackage{booktabs} 
\usepackage{slashed}
\usepackage{hyperref}

\renewcommand\sout{\bgroup \color{red} \ULdepth=-.5ex \ULset}

\begin{document}
\preprint{INHA-NTG-03/2025}
\title{\Large Production mechanism of hidden-charm pentaquark states
  $P_{c\bar{c}s}$ with strangeness $S=-1$ } 

\author{Samson Clymton}
\email[E-mail: ]{samson.clymton@apctp.org}
\affiliation{Department of Physics, Inha University,
Incheon 22212, Republic of Korea}
\affiliation{Asia Pacific Center for Theoretical Physics (APCTP),
  Pohang, Gyeongbuk 37673, Republic of Korea} 

\author{Hyun-Chul Kim}
\email[E-mail: ]{hchkim@inha.ac.kr}
\affiliation{Department of Physics, Inha University,
Incheon 22212, Republic of Korea}
\affiliation{School of Physics, Korea Institute for Advanced Study 
  (KIAS), Seoul 02455, Republic of Korea}

\author{Terry Mart}
\email[E-mail: ]{terry.mart@sci.ui.ac.id}
\affiliation{Departemen Fisika, FMIPA, Universitas Indonesia, Depok
  16424, Indonesia} 
\date{March, 2025}
\begin{abstract}
We investigate the hidden-charm pentaquark states with strangeness
$S=-1$ ($P_{c\bar{c}s}$) within an off-shell coupled-channel approach
based on effective Lagrangians that respect heavy-quark spin symmetry,
SU(3) flavor symmetry, and hidden local symmetry. All relevant
meson–baryon two-body channels composed of low-lying anti-charmed
mesons and singly-charmed baryons with $S=-1$, as well as the
$J/\psi\Lambda$ channel, are included. We find a total of eleven
negative-parity states and three positive-parity states. Among the
negative-parity states, the $P_{c\bar{c}s}(4338)$ and
$P_{c\bar{c}s}(4459)$ can possibly be interpreted as $\bar{D}\Xi_{c}$
and $\bar{D}^{*}\Xi_{c}$ molecular states, respectively. 
We identify a second state, $P_{c\bar{c}s}(4472)$,
located close to the $P_{c\bar{c}s}(4459)$ but with different spin and
width, which may correspond to the structure observed by the Belle
Collaboration. Both states are generated from the $\bar{D}^{*}\Xi_{c}$
channel and can be interpreted as spin partners. Their properties are
consistent with recent experimental observations, providing strong
support for the molecular interpretation of the $P_{c\bar{c}s}$
states. We also observe a two-pole structure near the
$\bar{D}_{s}^{*}\Lambda_{c}$ and $\bar{D}\Xi_{c}^{'}$ thresholds, and find
virtual and resonance states in the $\bar{D}^{*}\Xi_{c}^{'}$ channel
depending on spin-parity. 
\end{abstract}
\maketitle

\section{Introduction}
Since the observation of five hidden-charm pentaquark
states~\cite{LHCb:2015yax, LHCb:2019kea, LHCb:2021chn}, there has been
a great deal of experimental and theoretical work on these heavy
pentaquark states. Subsequently, the LHCb Collaboration reported the
existence of a neutral hidden-charm pentaquark with strangeness,
denoted as $P_{c\bar{c}s}(4459)$. The initial observation was made in
the $J/\psi\Lambda$ invariant mass spectrum from the $\Xi_b^- \to
J/\psi \Lambda K^-$ decay, revealing the $P_{c\bar{c}s}(4459)$ state
with mass $M_{P_{c\bar{c}s}} = (4458.8 \pm 2.9^{+4.7}_{-1.1})$
MeV/$c^2$ and width $\Gamma = (17.3 \pm 6.5^{+8.0}_{-5.7})$
MeV~\cite{LHCb:2021chn}. A second state, $P_{c\bar{c}s}(4338)$, was
identified in the $J/\psi\Lambda$ invariant mass spectrum from the
$B^- \to J/\psi \Lambda \bar{p}$ decay, with mass $M_{P_{c\bar{c}s}} =
(4338.2 \pm 0.7 \pm 0.4)$ MeV/$c^2$ and width $\Gamma = (7.0 \pm 1.2
\pm 1.3)$ MeV~\cite{LHCb:2022ogu}. Its spin-parity quantum numbers
were successfully determined to be $J^P=1/2^-$. These observations provide
compelling evidence for the existence of pentaquark states. Very
recently, the Belle Collaboration confirmed the existence of
$P_{c\bar{c}s}(4459)$, but reported a slightly larger mass:
$M_{P_{c\bar{c}s}} = (4471.7 \pm 4.8 \pm 0.6)$ MeV/$c^2$ and decay
width $\Gamma = (21.9 \pm 13.1 \pm 2.7)$
MeV~\cite{Belle:2025pey}. Considering that the CMS and LHCb
Collaborations have respectively reported the measurements of 
$\Lambda_b^0 \to J/\psi \Xi^- K^+$ and $\Xi_b^0 \to J/\psi \Xi^-
\pi^+$ decays, one may anticipate the possible existence of $S=-2$
hidden-charm pentaquark states in near future, denoted as
$P_{c\bar{c}ss}$~\cite{CMS:2024vnm, LHCb:2025lhk}. 

Before the discovery of hidden-charm pentaquarks with strangeness $S =
-1$, they had been predicted as molecular states composed of a heavy
meson and a singly heavy baryon system~\cite{Wu:2010jy, Xiao:2019gjd,
  Wang:2019nvm}. This interpretation was supported by experimental
observations, in which the mass of the $P_{c\bar{c}s}$ state was found
to be lower than the threshold energy of certain heavy meson–heavy
baryon systems. Moreover, the molecular nature of these states has
been extensively investigated within various theoretical
frameworks~\cite{Du:2021bgb, Karliner:2022erb, Giachino:2022pws,
  Wang:2022mxy, Yan:2022wuz, Zhu:2022wpi}. However, alternative
interpretations of the $P_{c\bar{c}s}$ states have also been proposed,
including compact pentaquark configurations~\cite{Shi:2021wyt,
  Li:2023aui, Zhang:2023teh}, or even kinematical
effects~\cite{Burns:2022uha}. Thus, it is of great importance to
investigate and understand the nature of the newly observed
$P_{c\bar{c}s}$ states within various theoretical frameworks. 

In our previous work, we successfully studied the production of
non-strange hidden-charm pentaquark states from meson–baryon
scattering using an off-shell coupled-channel
approach~\cite{Clymton:2024fbf}. We first constructed the kernel
Feynman amplitudes for the relevant channels based on an effective
Lagrangian that respects heavy-quark spin–flavor symmetry, hidden
local symmetry, and chiral symmetry. We then solved the
coupled-channel scattering integral equation involving seven different
coupled channels. By searching for poles corresponding to hidden-charm
pentaquark states in the complex energy plane, we found four
hidden-charm pentaquark states with negative parity, which were
associated with the $P_{c\bar{c}}$ states observed by the LHCb
Collaboration: $P_{c\bar{c}}(4312)$, $P_{c\bar{c}}(4380)$,
$P_{c\bar{c}}(4440)$, and $P_{c\bar{c}}(4457)$. In addition, we
predicted the existence of two further negative-parity and two
positive-parity $P_{c\bar{c}}$ states. 

Moreover, we provided an explanation for the absence of a
$P_{c\bar{c}}$ signal in the $J/\psi N$ photoproduction reported by
the GlueX experiment~\cite{GlueX:2019mkq}: destructive interference in
the $J/\psi N$ scattering, combined with suppression due to a dominant
positive-parity contribution, leads to the weakening of hidden-charm
pentaquark signals in the $J/\psi N$ channel~\cite{Clymton:2024fbf}. 

\begin{figure}[htp]
  \centering
  \includegraphics[scale=0.35]{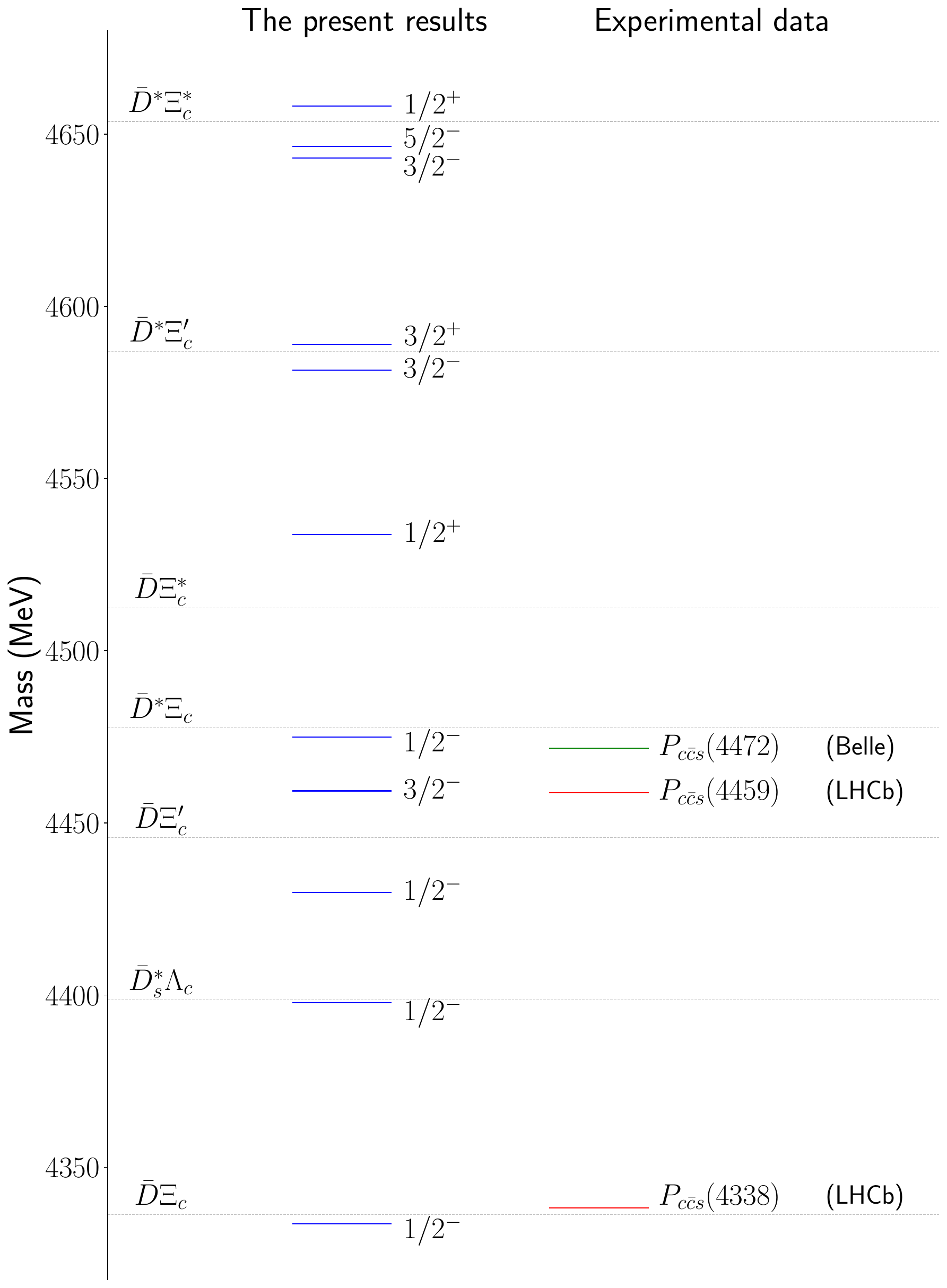}
  \caption{The mass spectrum of the $P_{c\bar{c}s}$'s, obtained from
    the present work. The experimental data are taken from the
    LHCb~\cite{LHCb:2021chn, LHCb:2022ogu} 
    and Belle~\cite{Belle:2025pey} measurements, respectively.} 
  \label{fig:1}  
\end{figure}

In the present work, we extend this approach to investigate
hidden-charm pentaquark states with strangeness $S = -1$, introducing
the relevant heavy meson–heavy baryon scattering channels. As a
result, we have found eight resonances with negative parity and three
with positive parity for spin $1/2$, $3/2$, and $5/2$. Two of these
can be associated with the experimentally observed $P_{c\bar{c}s}$ states:
$P_{c\bar{c}s}(4338)$ and $P_{c\bar{c}s}(4459)$. 
Figure~\ref{fig:1} summarizes the results for the predictions of the
$P_{c\bar{c}s}$ pentaquarks obtained from the current work. The
remaining nine pentaquark states are considered predictions. We want
to emphasize that the \emph{two} hidden-charm pentaquark states below
the $\bar{D}^* \Xi_c$ threshold should be different ones with
different spins.  

The present work is organized as follows: In Sec.\ref{sec:2}, we
present the theoretical formalism employed to investigate hidden-charm
strange pentaquarks, incorporating all possible two-body channels
composed of ground-state charmed baryons and anti-charmed mesons,
along with the additional $J/\psi\Lambda$ channel. The resulting
scattering matrix is analyzed by examining its behavior in both the
real and complex energy domains. In Sec.\ref{sec:3}, we discuss our
findings, with particular focus on the molecular nature of each
resonant state. Finally, we conclude the paper with a summary and
concluding remarks in Sec.~\ref{sec:4}. 

\section{Coupled-channel formalism\label{sec:2}} 
\begin{figure}[htbp]
  \centering
  \includegraphics[scale=1.0]{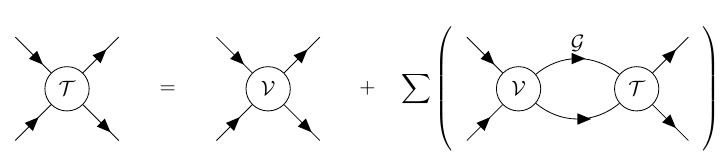}
  \caption{Graphical representation of the coupled integral scattering
    equation with the two-body intermediate states.}   
  \label{fig:2}
\end{figure}

The scattering amplitude is defined as 
\begin{align}
\mathcal{S}_{fi} = \delta_{fi} - i (2\pi)^4 \delta(P_f - P_i)
  \mathcal{T}_{fi}, 
\end{align}
where $P_i$ and $P_f$ denote the total four-momenta of the initial and
final states, respectively. The transition amplitudes
$\mathcal{T}_{fi}$ are obtained from the Bethe-Salpeter integral
equation with the two-body Feynman kernel amplitudes  
\begin{align}
\mathcal{T}_{fi} (p',p;s) =\, \mathcal{V}_{fi}(p',p;s) 
+ \frac{1}{(2\pi)^4}\sum_k \int d^4q 
\mathcal{V}_{fk}(p',q;s)\mathcal{G}_{k}(q;s) \mathcal{T}_{ki}(q,p;s),  
\label{eq:2}
\end{align}
where $p$ and $p'$ indicate the relative four-momenta of the initial
and final states, respectively. $q$ represents the off-mass-shell
momentum for the intermediate states in the center of mass (CM)
frame. $s$ is the square of the total energy, which is one of the
Mandelstam variables, $s=P_i^2=P_f^2$.
The coupled integral equations presented in Eq.~\eqref{eq:2} can be
visualized as shown in Fig.~\ref{fig:2}. 

To reduce the complexity of the four-dimensional integral equations,
we implement a three-dimensional reduction. Among various methods for
three-dimensional reduction, we utilize the Blankenbecler-Sugar
formalism~\cite{Blankenbecler:1965gx, Aaron:1968aoz}, which expresses
the two-body propagator in the form of  
\begin{align}
  \mathcal{G}_k(q) =\;
  \delta\left(q_0-\frac{E_{k1}(\bm{q})-E_{k2}(\bm{q})}{2}\right)
  \frac{\pi}{E_{k1}(\bm{q})E_{k2}(\bm{q})}
  \frac{E_k(\bm{q})}{s-E_k^2(\bm{q})},  
\label{eq:4}
\end{align}
where $E_k$ is the total on-mass-shell energy of the
intermediate state, $E_k = E_{k1}+E_{k2}$, and $\bm{q}$ denotes the
three-momentum of the intermediate state. Note that the spinor
contributions from the meson-baryon propagator $G_k$ have been
incorporated into the matrix elements of $\mathcal{V}$ and
$\mathcal{T}$. By applying Eq.~\eqref{eq:4}, we derive the following
coupled integral equations    
\begin{align}
  \mathcal{T}_{fi} (\bm{p}',\bm{p}) =\, \mathcal{V}_{fi}
  (\bm{p}',\bm{p}) 
  +\frac{1}{(2\pi)^3}\sum_k\int \frac{d^3q}{2E_{k1}(\bm{q})E_{k2}
  (\bm{q})} \mathcal{V}_{fk}(\bm{p}',\bm{q})\frac{E_k
  (\bm{q})}{s-E_k^2(\bm{q})+i\varepsilon} 
  \mathcal{T}_{ki}(\bm{q},\bm{p}),
  \label{eq:BS-3d}
\end{align}
where $\bm{p}$ and $\bm{p}'$ are the relative three-momenta of the
initial and final states in the CM frame, respectively. 

\begin{figure}[htbp]
  \centering
  \includegraphics[scale=0.35]{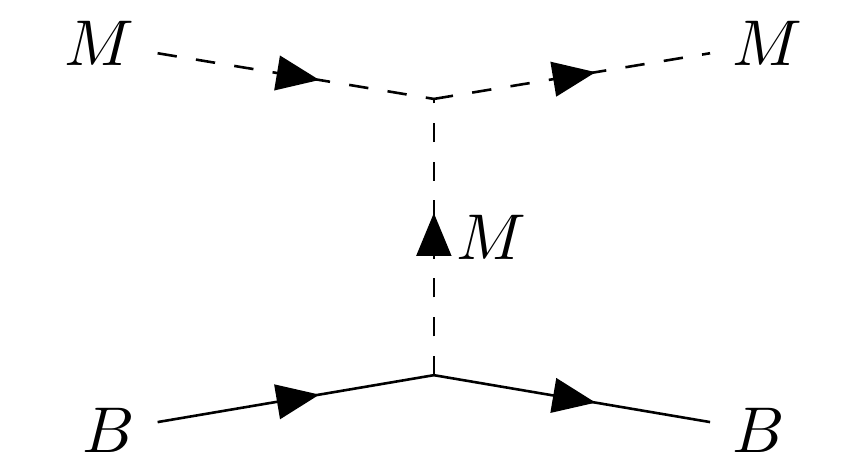}
  \caption{$t$-channel diagrams for the
          meson-exchanged diagrams. $M$ and $B$ stand for the
          meson and baryon, respectively.} 
  \label{fig:3}
\end{figure}

To investigate the dynamical generation of the pentaquark states with
strangeness $S = -1$, we construct two-body coupled channels by
combining the charmed meson triplet with the singly charmed baryon
antitriplet and sextet, selecting only the combinations that yield
strangeness $S = -1$. In addition, we include the $J/\psi \Lambda$
channel, as the $P_{c\bar{c}s}$ states have been experimentally
observed to decay into $J/\psi \Lambda$. This results 
in nine distinct channels with $S = -1$: $J/\psi \Lambda$, 
$\bar{D}_s\Lambda_c$, $\bar{D}\Xi_c$, $\bar{D}_s^*\Lambda_c$,
$\bar{D}\Xi_c'$, $\bar{D}^*\Xi_c$, $\bar{D}\Xi_c^*$, $\bar{D}^*\Xi_c'$
and $\bar{D}^*\Xi_c^*$. We construct the kernel matrix using one-meson
exchange tree-level diagrams, as illustrated in Fig.~\ref{fig:3}.
In our approach, we exclude any pole diagrams in the $s$-channel,
since we want to demonstrate how the hidden-charm pentaquark states are
generated dynamically by the interplay between various coupled
channels. Nevertheless, we want to mention the importance of the
$s$-channel pole diagrams in describing existing resonances. There
will be certain contributions from them, so that the pole positions
and widths of the $P_{c\bar{c}s}$'s may be influenced by
them. However, it is of great difficult to consider them without
additional uncertainties involved, in particular, when the predicted
$P_{c\bar{c}s}$'s are concerned. 
 
Thus, our primary focus is on the $t$-channel diagrams, which play a
crucial role in dynamically generating the $P_{c\bar{c}s}$ states. The 
$u$-channel diagrams contain the exchange of doubly-charmed baryons
with masses around 3.5 GeV and are significantly suppressed in
magnitude compared to the $t$-channel diagrams. As a result, their
contributions are negligible, and we therefore omit them from our
analysis. 

The vertex interactions are determined by an effective Lagrangian that
respects heavy-quark spin symmetry, hidden local symmetry, and
flavor SU(3) symmetry~\cite{Casalbuoni:1996pg}. The mesonic vertices
are computed using the effective Lagrangian given by 
\begin{align}
  \mathcal{L}_{PP\mathbb{V}} &= -i\frac{\beta g_V}{\sqrt{2}}\,
  P^{\dagger}_a  \overleftrightarrow{\partial_\mu} P_b\,
   \mathbb{V}^\mu_{ba} +i\frac{\beta g_V}{\sqrt{2}}\,P'^{\dagger}_a
                               \overleftrightarrow{\partial_\mu}
                               P'_b\, \mathbb{V}^\mu_{ab},\\ 
    \mathcal{L}_{PP\sigma} &= -2g_\sigma M P^\dagger_a \sigma P_a
  -2g_\sigma M P'^\dagger_a\sigma P'_a ,\\     %
    \mathcal{L}_{P^*P^*\mathbb{P}} &= -\frac{g}{f_\pi}
  \epsilon^{\mu\nu\alpha\beta}P^{*\dagger}_{a\nu}\,
   \overleftrightarrow{\partial_\mu}\,
P^*_{b\beta}\partial_\alpha \mathbb{P}_{ba} -\frac{g}{f_\pi}
  \epsilon^{\mu\nu\alpha\beta}  P'^{*\dagger}_{a\nu}\,
 \overleftrightarrow{\partial_\mu}\,
 P'^*_{b\beta}\partial_\alpha
                                     \mathbb{P}_{ab} ,\\     %
    \mathcal{L}_{P^*P^*\mathbb{V}} & = i\frac{\beta g_V}{\sqrt{2}} \,
 P^{*\dagger}_{a\nu}  \overleftrightarrow{\partial_\mu}
 P^{*\nu}_b \mathbb{V}_{ba}^\mu + i2\sqrt{2} \lambda g_VM^*
 P^{*\dagger}_{a\mu}  P^*_{b\nu}\mathbb{V}_{ba}^{\mu\nu}\cr
  &\;\;\;\;-i\frac{\beta g_V}{\sqrt{2}} \,P'^{*\dagger}_{a\nu}
    \overleftrightarrow{\partial_\mu} P'^{*\nu}_b
    \mathbb{V}_{ab}^\mu-i2\sqrt{2}\lambda g_VM^*
    P'^{*\dagger}_{a\mu} P'^*_{b\nu}\mathbb{V}_{ab}^{\mu\nu} ,\\
  \mathcal{L}_{P^*P^*\sigma} &= 2g_\sigma M^* P^{*\dagger}_{a\mu}
\sigma P^{*\mu}_a+2g_\sigma M^*
 P'^{*\dagger}_{a\mu}\sigma P'^{*\mu}_a ,\\    %
    \mathcal{L}_{P^*P\mathbb{P}} &= -\frac{2g}{f_\pi}  \sqrt{MM^*}\,
      \left( P^{\dagger}_a P^*_{b\mu}+P^{*\dagger}_{a\mu} P_b\right)\,
 \partial^\mu \mathbb{P}_{ba}+\frac{2g}{f_\pi} \sqrt{MM^*}\,
    \left(P'^{\dagger}_a P'^*_{b\mu}+P'^{*\dagger}_{a\mu} P'_b\right)\,
  \partial^\mu \mathbb{P}_{ab},\\
  \mathcal{L}_{P^*P\mathbb{V}} &= -i\sqrt{2}\lambda g_V\,
\epsilon^{\beta\alpha\mu\nu} \left(P^{\dagger}_a
\overleftrightarrow{\partial_\beta} P^*_{b\alpha} +
P^{*\dagger}_{a\alpha}  \overleftrightarrow{\partial_\beta}
  P_b\right)\,\left(\partial_\mu\mathbb{V}_{\nu}\right)_{ba}\cr 
 &\;\;\;\;-i\sqrt{2}\lambda g_V\, \epsilon^{\beta\alpha\mu\nu}
   \left(P'^{\dagger}_a\overleftrightarrow{\partial_\beta}
   P'^*_{b\alpha}+P'^{*\dagger}_{a\alpha}
   \overleftrightarrow{\partial_\beta}P'_b\right)\,
   \left(\partial_\mu\mathbb{V}_{\nu}\right)_{ab}.
\end{align}
where $\overleftrightarrow{\partial} =
\overrightarrow{\partial}-\overleftarrow{\partial}$. The symbol $\sigma$ represents the lowest isoscalar-scalar meson. The matrices for heavy mesons and anti-heavy mesons $P^{(*)}$ and $P'^{(*)}$ are defined as 
\begin{align}
  P = \left(D^0,D^+,D_s^+\right), \hspace{0.5 cm}
  P^*_\mu =\left(D^{*0}_\mu,D^{*+}_\mu,D_{s\mu}^{*+}\right),
  \hspace{0.5 cm} P' =(\bar{D}^0,\,D^-,\,D_s^-),
  \hspace{0.5 cm} P'^*_\mu =(\bar{D}^{*0}_\mu,\,D^{*-}_\mu,\,D^{*-}_{s\mu}),
\end{align}
while the matrices for light pseudoscalar and vector mesons are expressed as
\begin{align}
    \mathbb{P} = 
    \begin{pmatrix}
        \frac{1}{\sqrt{2}} \pi^0+\frac{1}{\sqrt{6}}\eta & \pi^+ & K^+\\
        \pi^- & -\frac{1}{\sqrt{2}} \pi^0+\frac{1}{\sqrt{6}}\eta & K^0\\
        K^- & \bar{K}^0 & -\frac{2}{\sqrt{6}}\eta
    \end{pmatrix},\;\;\;\;
    \mathbb{V}_\mu = \begin{pmatrix}
        \frac{1}{\sqrt{2}} \rho^0_\mu+\frac{1}{\sqrt{2}}\omega_\mu &
        \rho_\mu^+ & K_\mu^{*+}\\
        \rho_\mu^- & -\frac{1}{\sqrt{2}} \rho_\mu^0+\frac{1}{\sqrt{2}}
        \omega_\mu & K_\mu^{*0} \\
        K_\mu^{*-} & \bar{K}^{*0}_\mu & \phi_\mu
    \end{pmatrix}.
\end{align}

The coupling constants in the Lagrangian are taken from
Ref.~\cite{Isola:2003fh}: $g = 0.59 \pm 0.07 \pm 0.01$, determined
from experimental measurements of the full width of the $D^{*+}$; $g_V
= m_\rho / f_\pi \approx 5.8$, obtained via the
Kawarabayashi-Suzuki-Riazuddin-Fayyazuddin (KSRF) relation with
$f_\pi = 132~\mathrm{MeV}$; $\beta \approx 
0.9$~\cite{Kawarabayashi:1966kd, Riazuddin:1966sw}, based on the 
vector meson dominance in the radiative decay of heavy
mesons; and $\lambda = -0.56~\mathrm{GeV}^{-1}$, derived from
light-cone sum rules and lattice QCD results. It should be noted that
we adopt a different sign convention for $\lambda$ compared to
Ref.~\cite{Isola:2003fh}, as we use the same phase for heavy vector
mesons as in Ref.~\cite{Casalbuoni:1996pg}. 
The coupling constant for the sigma meson is used to evaluate the
$2\pi$ transition of $D_s(1^+)$ in Ref.~\cite{Bardeen:2003kt}. The
coupling for the lowest isoscalar-scalar meson is given by 
$g_\sigma = \frac{g_\pi}{2\sqrt{6}}$ with $g_\pi = 3.73$.

Regarding the effective Lagrangian for the heavy baryon, we adopt it
from Ref.~\cite{Liu:2011xc}, which considers a more comprehensive form
of the Lagrangian as discussed in Ref.~\cite{Yan:1992gz}. The baryonic
interaction vertices in the tree-level meson-exchange diagrams are
characterized by the following effective Lagrangian: 
\begin{align}
    \mathcal{L}_{B_{\bar{3}}B_{\bar{3}}\mathbb{V}}& =
 \frac{i\beta_{\bar{3}}g_V}{2\sqrt{2}M_{\bar{3}}}
  \left(\bar{B}_{\bar{3}}\overleftrightarrow{\partial_\mu}
  \mathbb{V}^\mu B_{\bar{3}}\right) ,\\
  \mathcal{L}_{B_{\bar{3}}B_{\bar{3}}\sigma}&=l_{\bar{3}}
 \left(\bar{B}_{\bar{3}}\sigma B_{\bar{3}}\right) ,\\ 
    \mathcal{L}_{B_6B_6\mathbb{P}}&= i\frac{g_1}{2f_\pi M_6}\bar{B}_{6}\gamma_5\left(\gamma^\alpha\gamma^\beta-g^{\alpha\beta}\right)\overleftrightarrow{\partial_\alpha}\partial_\beta\mathbb{P} B_{6} ,\\
    \mathcal{L}_{B_6B_6\mathbb{V}}&= -i\frac{\beta_6 g_V}{2\sqrt{2}M_6}\left(\bar{B}_{6}\overleftrightarrow{\partial_\alpha}\mathbb{V}^\alpha B_{6}\right)-\frac{i\lambda_6g_V}{3\sqrt{2}}\left(\bar{B}_{6}\gamma_\mu\gamma_\nu \mathbb{V}^{\mu\nu}B_{6}\right) ,\\
    \mathcal{L}_{B_6B_6\sigma}&= -l_6\left(\bar{B}_{6}\sigma B_6\right) ,\\
    \mathcal{L}_{B_6^*B_6^*\mathbb{P}}&= \frac{3g_1}{4f_\pi M_6^*}\epsilon^{\mu\nu\alpha\beta}\left(\bar{B}_{6\mu}^*\overleftrightarrow{\partial_\nu}\partial_\alpha\mathbb{P} B_{6\beta}^*\right) ,\\
    \mathcal{L}_{B_6^*B_6^*\mathbb{V}}&= i\frac{\beta_6 g_V}{2\sqrt{2}M_6^*}\left(\bar{B}_{6\mu}^*\overleftrightarrow{\partial_\alpha}\mathbb{V}^\alpha B_{6}^{*\mu}\right)+\frac{i\lambda_6g_V}{\sqrt{2}} \left(\bar{B}^*_{6\mu} \mathbb{V}^{\mu\nu}B^*_{6\nu}\right) ,\\
    \mathcal{L}_{B_6^*B_6^*\sigma}&= l_6\left(\bar{B}^*_{6\mu}\sigma B^{*\mu}_6\right) ,\\
    \mathcal{L}_{B_6B_6^*\mathbb{P}}&= \frac{g_1}{4f_\pi}\sqrt{\frac{3}{M_6^*M_6}}\epsilon^{\mu\nu\alpha\beta}\left[\left(\bar{B}_{6}\gamma_5\gamma_\mu\overleftrightarrow{\partial_\nu} \partial_\alpha\mathbb{P} B_{6\beta}^*\right)+\left(\bar{B}_{6\mu}^*\gamma_5\gamma_\nu\overleftrightarrow{\partial_\alpha} \partial_\beta\mathbb{P}B_6 \right)\right] ,\\
    \mathcal{L}_{B_6B_6^*\mathbb{V}}&= \frac{i\lambda_6g_V}{\sqrt{6}}\left[\bar{B}_{6}\gamma_5\left(\gamma_\mu +\frac{i\overleftrightarrow{\partial_\mu}}{2\sqrt{M_6^*M_6}}\right) \mathbb{V}^{\mu\nu}B^*_{6\nu}+\bar{B}_{6\mu}^*\gamma_5\left(\gamma_\nu -\frac{i\overleftrightarrow{\partial_\nu}}{2\sqrt{M_6^*M_6}}\right) \mathbb{V}^{\mu\nu}B_{6}\right] ,\\
    \mathcal{L}_{B_6B_{\bar{3}}\mathbb{P}}&= -\frac{g_4}{\sqrt{3}f_\pi}\left[\bar{B}_6\gamma_5\left(\gamma_\mu +\frac{i\overleftrightarrow{\partial_\mu}}{2\sqrt{M_6M_{\bar{3}}}} \right) \partial^\mu\mathbb{P} B_{\bar{3}}+\bar{B}_{\bar{3}}\gamma_5\left(\gamma_\mu -\frac{i\overleftrightarrow{\partial_\mu}}{2\sqrt{M_6M_{\bar{3}}}} \right) \partial^\mu\mathbb{P}\,B_{6}\right] ,\\
    \mathcal{L}_{B_6B_{\bar{3}}\mathbb{V}}&= i\frac{\lambda_{6\bar{3}}\,g_V}{\sqrt{6M_6M_{\bar{3}}}}\epsilon^{\mu\nu\alpha\beta}\left[\left(\bar{B}_{6}\gamma_5\gamma_\mu \overleftrightarrow{\partial_\nu}\partial_\alpha\mathbb{V}_{\beta} B_{\bar{3}}\right)+\left(\bar{B}_{\bar{3}}\gamma_5\gamma_\mu \overleftrightarrow{\partial_\nu}\partial_\alpha\mathbb{V}_{\beta} B_{6}\right)\right] ,\\
    \mathcal{L}_{B_6^*B_{\bar{3}}\mathbb{P}}&= -\frac{g_4}{f_\pi}\left[\left(\bar{B}^*_{6\mu}\partial^\mu\mathbb{P} B_{\bar{3}}\right)+\left(\bar{B}_{\bar{3}}\partial^\mu\mathbb{P} B^*_{6\mu}\right)\right] ,\\
    \mathcal{L}_{B_6^*B_{\bar{3}}\mathbb{V}}&= i\frac{\lambda_{6\bar{3}}\,g_V}{\sqrt{2M_6^*M_{\bar{3}}}}\epsilon^{\mu\nu\alpha\beta}\left[\left(\bar{B}^*_{6\mu}\overleftrightarrow{\partial_\nu}\partial_\alpha\mathbb{V}_{\beta} B_{\bar{3}}\right)+\left(\bar{B}_{\bar{3}}\overleftrightarrow{\partial_\nu}\partial_\alpha\mathbb{V}_{\beta} B^*_{6\mu}\right)\right] ,
\end{align}
where the heavy baryon fields are expressed as
\begin{align}
    &B_{\bar{3}}=
    \begin{pmatrix}
        0 & \Lambda_c^+ & \Xi_c^+\\
        -\Lambda_c^+ & 0 & \Xi_c^0\\
        -\Xi_c^+ & -\Xi_c^0 & 0
    \end{pmatrix},\;\;
    B_6 =
    \begin{pmatrix}
        \Sigma_c^{++} & \frac{1}{\sqrt{2}}\Sigma_c^+ & \frac{1}{\sqrt{2}}\Xi{'}_c^+\\
        \frac{1}{\sqrt{2}}\Sigma_c^+ & \Sigma_c^0 & \frac{1}{\sqrt{2}}\Xi{'}_c^0\\
        \frac{1}{\sqrt{2}}\Xi{'}_c^+ & \frac{1}{\sqrt{2}}\Xi{'}_c^0 & \Omega_c^0
    \end{pmatrix},\;\;
    B_6^* =
    \begin{pmatrix}
        \Sigma_c^{*++} & \frac{1}{\sqrt{2}}\Sigma_c^{*+} & \frac{1}{\sqrt{2}}\Xi_c^{*+}\\
        \frac{1}{\sqrt{2}}\Sigma_c^{*+} & \Sigma_c^{*0} & \frac{1}{\sqrt{2}}\Xi_c^{*0}\\
        \frac{1}{\sqrt{2}}\Xi_c^{*+} & \frac{1}{\sqrt{2}}\Xi_c^{*0} & \Omega_c^{*0}
    \end{pmatrix}.
\end{align}
Here, $B_\mu$ represents the spin 3/2 Rarita-Schwinger field, which must satisfy the following constraints
\begin{align}
  p^\mu B_\mu = 0 \hspace{0.5 cm}{\rm and}\hspace{0.5 cm}
  \gamma^\mu B_\mu = 0 .
\end{align}
The coupling constants within the effective Lagrangian are specified
as follows~\cite{Liu:2011xc,Chen:2019asm}: $\beta_{\bar{3}} = 6/g_V$,
$\beta_6=-2\beta_{\bar{3}}$, $\lambda_6=-3.31\,\mathrm{GeV}^{-1}$,
$\lambda_{6\bar{3}}=-\lambda_6/\sqrt{8}$, $g_1=0.942$,
$g_4=0.999$, $l_{\bar{3}}= -3.1$ and $l_6=-2l_{\bar{3}}$.
The sign conventions employed here are consistent with
those in Refs.~\cite{Chen:2019asm, Dong:2021juy}. 

For the inclusion of hidden-charm channels, we require an effective
Lagrangian that describes the coupling between heavy mesons and
quarkonium. We utilize the Lagrangian from
Ref.~\cite{Colangelo:2003sa}, given by  
\begin{align}
    \mathcal{L}_{PPJ/\psi} &= -ig_\psi M\sqrt{m_{J}}\left(J/\psi^\mu P^\dagger\overleftrightarrow{\partial_\mu}P{'}^{\dagger}\right) + \mathrm{h.c.,}\\
    \mathcal{L}_{P^*PJ/\psi} &= ig_\psi\sqrt{\frac{MM^*}{m_{J}}}\epsilon^{\mu\nu\alpha\beta} \partial_\mu J/\psi_\nu\left(P^\dagger\overleftrightarrow{\partial_\alpha}P^*{'}^\dagger_\beta+P_{\beta}^{*\dagger}\overleftrightarrow{\partial_\alpha}P{'}^{\dagger}\right)+\mathrm{h.c.,}\\
    \mathcal{L}_{P^*P^*J/\psi} &= ig_\psi M^*\sqrt{m_J}(g^{\mu\nu}g^{\alpha\beta}-g^{\mu\alpha}g^{\nu\beta}+g^{\mu\beta}g^{\nu\alpha}) \left(J/\psi_\mu P_{\nu}^{*\dagger}\overleftrightarrow{\partial_\alpha}P^*{'}^\dagger_\beta\right)+\mathrm{h.c.}
\end{align}

In the present study, we focus exclusively on vector quarkonia due to
their direct experimental relevance. Nevertheless, extending the
analysis to include pseudoscalar states is straightforward, as we
apply heavy quark spin symmetry to the quarkonium sector as
well~\cite{Casalbuoni:1992fd}. In the absence of experimental data for
the $J/\psi \to D\bar{D}$ decay, Shimizu et al.~\cite{Shimizu:2017xrg}
provided an estimate for the coupling constant $g_\psi$ using the
following approach: they first determined the coupling constant
$g_{\phi K\bar{K}}$ from the experimental decay width of $\phi \to
K\bar{K}$. Assuming that the decay mechanisms of the $J/\psi$ are
analogous to those of the $\phi$, aside from mass differences, they
estimated the coupling constant as $g_\psi =
0.679\,\mathrm{GeV}^{-3/2}$. 
The coupling constants between heavy baryons and heavy mesons are 
formulated following Ref.~\cite{Shimizu:2017xrg}: 
\begin{align}
    \mathcal{L}_{B_8B_3P} &=
   -g_{I{\bar{3}}}\sqrt{M}\bar{B}_{\bar{3}}\gamma_5
   P N+\mathrm{h.c.},\\ 
    \mathcal{L}_{B_8B_3P^*} &=
    g_{I{\bar{3}}}\sqrt{M^*}\bar{B}_{\bar{3}}\gamma^\mu
    P_\mu^* N+\mathrm{h.c.,}\\ 
    \mathcal{L}_{B_8B_6P}&= g_{I6} \sqrt{3M} \bar{B}_{6}\gamma_5 B_8 P
        + \mathrm{h.c.,}\\ 
    \mathcal{L}_{B_8B_6P^*}&= g_{I6}\sqrt{\frac{M^*}{3}}
   \bar{B}_{6}\gamma^\nu B_8 P^*_\nu +
   \mathrm{h.c.,}\\ 
    \mathcal{L}_{B_8B_6^*P^*}&= 2 g_{I6}\sqrt{M^*}
    \bar{B}_{6}^\mu\gamma_5 B_8 P^*_\mu +
    \mathrm{h.c.}. 
\end{align}

We adopt the coupling constants $g_{I\bar{3}} =
-9.88\,\mathrm{GeV}^{-1/2}$ and $g_{I6} = 1.14\,\mathrm{GeV}^{-1/2}$
from Ref.~\cite{Shimizu:2017xrg}. It should be emphasized that the
coupling to hidden-charm channels has only a minimal impact on the
resonance production mechanism. 
The calculations indicate that, although these coupling constants are
based on approximate estimates, the predicted masses of the
hidden-charm pentaquarks remain largely unchanged. 
This finding suggests that the $J/\psi \Lambda$ channel
contributes only marginally to the formation of heavy pentaquarks. 
 
\begin{table}[htbp]
  \caption{\label{tab:1} Values of the IS factors and
          $\Lambda-m$ for the corresponding $t$-channel diagrams for
          the given reactions. $\Lambda$ denotes the cutoff mass
          and $m$ stands for the mass of the exchanged particle, given
          in units of MeV.  
        } 
   \renewcommand{\arraystretch}{1.2}
  \begin{ruledtabular}
  \centering\begin{tabular}{lccr}
   \multirow{2}{*}{Reactions} & \multirow{2}{*}{Exchange particles} & 
   \multirow{2}{*}{IS} &   \multirow{2}{*}{$\Lambda-m$} \\
   & & & 
   \\
   \hline\\[-2.5ex]
   $J/\psi\Lambda\to\bar{D}_s\Lambda_c$ 
   & $\bar{D}_s$,$\bar{D}_s^*$  & $-\frac{1}{3}\sqrt{6}$ & $600$ \\
   $J/\psi\Lambda\to\bar{D}\Xi_c$ 
   & $\bar{D}$,$\bar{D}^*$  & $-\frac{1}{3}\sqrt{3}$ & $600$ \\
   $J/\psi\Lambda\to\bar{D}_s^*\Lambda_c$ 
   & $\bar{D}_s$,$\bar{D}_s^*$  & $-\frac{1}{3}\sqrt{6}$ & $600$ \\
   $J/\psi\Lambda\to\bar{D}\Xi_c'$ 
   & $\bar{D}$,$\bar{D}^*$ & $\frac{1}{2}\sqrt{6}$ & $600$ \\
   $J/\psi\Lambda\to\bar{D}^*\Xi_c$ 
   & $\bar{D}$,$\bar{D}^*$ & $-\frac{1}{3}\sqrt{3}$ & $600$ \\
   $J/\psi\Lambda\to\bar{D}\Xi_c^*$ 
   & $\bar{D}$,$\bar{D}^*$ & $\frac{1}{2}\sqrt{6}$ & $600$ \\
   $J/\psi\Lambda\to\bar{D}^*\Xi_c'$ 
   & $\bar{D}$,$\bar{D}^*$ & $\frac{1}{2}\sqrt{6}$ & $600$ \\
   $J/\psi\Lambda\to\bar{D}^*\Xi_c^*$ 
   & $\bar{D}$,$\bar{D}^*$ & $\frac{1}{2}\sqrt{6}$ & $600$ \\
   $\bar{D}_s\Lambda_c\to\bar{D}_s\Lambda_c$ 
   & $\sigma$  & $2$ & $600$ \\
   $\bar{D}_s\Lambda_c\to\bar{D}\Xi_c$ 
   & $K^*$  & $\sqrt{2}$ & $600$ \\
   $\bar{D}_s\Lambda_c\to\bar{D}\Xi_c'$ 
   & $K^*$ & $-1$ & $600$ \\
   $\bar{D}_s\Lambda_c\to\bar{D}^*\Xi_c$ 
   & $K^*$ & $\sqrt{2}$ & $600$ \\
   $\bar{D}_s\Lambda_c\to\bar{D}\Xi_c^*$ 
   & $K^*$ & $-1$ & $600$ \\
   $\bar{D}_s\Lambda_c\to\bar{D}^*\Xi_c'$ 
   & $K$,$K^*$ & $-1$ & $600$ \\
   $\bar{D}_s\Lambda_c\to\bar{D}^*\Xi_c^*$ 
   & $K$,$K^*$ & $-1$ & $600$ \\
   $\bar{D}\Xi_c\to\bar{D}\Xi_c$ 
   & $\rho$   & $-\frac{3}{2}$ & $600$ \\
   & $\omega$ & $\frac{1}{2}$ & $600$ \\
   & $\sigma$ & $2$ & $600$ \\
   $\bar{D}\Xi_c\to\bar{D}_s^*\Lambda_c$ 
   & $K^*$  & $\sqrt{2}$ & $600$ \\
   $\bar{D}\Xi_c\to\bar{D}\Xi_c'$ 
   & $\rho$   & $-\frac{3}{4}\sqrt{2}$ & $600$ \\
   & $\omega$ & $\frac{1}{4}\sqrt{2}$ & $600$ \\
   $\bar{D}\Xi_c\to\bar{D}^*\Xi_c$ 
   & $\rho$   & $-\frac{3}{2}$ & $600$ \\
   & $\omega$ & $\frac{1}{2}$ & $600$ \\
   $\bar{D}\Xi_c\to\bar{D}\Xi_c^*$ 
   & $\rho$   & $-\frac{3}{4}\sqrt{2}$ & $600$ \\
   & $\omega$ & $\frac{1}{4}\sqrt{2}$ & $600$ \\
   $\bar{D}\Xi_c\to\bar{D}^*\Xi_c'$ 
   & $\pi$    & $-\frac{3}{4}\sqrt{2}$ & $600$ \\
   & $\eta$   & $\frac{1}{4}\sqrt{2}$ & $600$ \\
   & $\rho$   & $-\frac{3}{4}\sqrt{2}$ & $600$ \\
   & $\omega$ & $\frac{1}{4}\sqrt{2}$ & $600$ \\
   $\bar{D}\Xi_c\to\bar{D}^*\Xi_c^*$ 
   & $\pi$    & $-\frac{3}{4}\sqrt{2}$ & $600$ \\
   & $\eta$   & $\frac{1}{4}\sqrt{2}$ & $600$ \\
   & $\rho$   & $-\frac{3}{4}\sqrt{2}$ & $600$ \\
   & $\omega$ & $\frac{1}{4}\sqrt{2}$ & $600$ \\
   $\bar{D}_s^*\Lambda_c\to\bar{D}_s^*\Lambda_c$ 
   & $\sigma$  & $2$ & $600$ \\
   $\bar{D}_s^*\Lambda_c\to\bar{D}\Xi_c'$ 
   & $K$,$K^*$ & $-1$ & $600$ \\
   $\bar{D}_s^*\Lambda_c\to\bar{D}^*\Xi_c$ 
   & $K^*$ & $\sqrt{2}$ & $600$ \\
   $\bar{D}_s^*\Lambda_c\to\bar{D}\Xi_c^*$ 
   & $K$,$K^*$ & $-1$ & $600$ \\
   $\bar{D}_s^*\Lambda_c\to\bar{D}^*\Xi_c'$ 
   & $K$,$K^*$ & $-1$ & $600$ \\
   $\bar{D}_s^*\Lambda_c\to\bar{D}^*\Xi_c^*$ 
   & $K$,$K^*$ & $-1$ & $600$ \\
   $\bar{D}\Xi_c'\to\bar{D}\Xi_c'$ 
   & $\rho$   & $-\frac{3}{4}$ & $600$ \\
   & $\omega$ & $\frac{1}{4}$ & $600$ \\
   & $\sigma$ & $1$ & $600$ \\
   $\bar{D}\Xi_c'\to\bar{D}^*\Xi_c$ 
   & $\pi$    & $-\frac{3}{4}\sqrt{2}$ & $600$ \\
   & $\eta$   & $\frac{1}{4}\sqrt{2}$ & $600$ \\
   & $\rho$   & $-\frac{3}{4}\sqrt{2}$ & $600$ \\
   & $\omega$ & $\frac{1}{4}\sqrt{2}$ & $600$ \\
   $\bar{D}\Xi_c'\to\bar{D}\Xi_c^*$ 
   & $\rho$   & $-\frac{3}{4}$ & $600$ \\
   & $\omega$ & $\frac{1}{4}$ & $600$ \\
  \end{tabular}
    \end{ruledtabular}
\end{table}
\begin{table}[htbp]
  \addtocounter{table}{-1}
  \caption{The values of the IS factors and
          $\Lambda-m$ for the corresponding $t$-channel diagrams for
          the given reactions. The $\Lambda$ denotes the cutoff mass
          and $m$ stands for the mass of the exchanged particle, given
          in units of MeV (continued).  
        } 
   \renewcommand{\arraystretch}{1.2}
  \begin{ruledtabular}
  \centering\begin{tabular}{lccr}
   \multirow{2}{*}{Reactions} & \multirow{2}{*}{Exchange particles} & 
   \multirow{2}{*}{IS} &   \multirow{2}{*}{$\Lambda-m$} \\
   & & & 
   \\
   \hline\\[-2.5ex]
   $\bar{D}\Xi_c'\to\bar{D}^*\Xi_c'$ 
   & $\pi$    & $-\frac{3}{4}$ & $600$ \\
   & $\eta$   & $-\frac{1}{12}$ & $600$ \\
   & $\rho$   & $-\frac{3}{4}$ & $600$ \\
   & $\omega$ & $\frac{1}{4}$ & $600$ \\
   $\bar{D}\Xi_c'\to\bar{D}^*\Xi_c^*$ 
   & $\pi$    & $-\frac{3}{4}$ & $600$ \\
   & $\eta$   & $-\frac{1}{12}$ & $600$ \\
   & $\rho$   & $-\frac{3}{4}$ & $600$ \\
   & $\omega$ & $\frac{1}{4}$ & $600$ \\
$\bar{D}^*\Xi_c\to\bar{D}^*\Xi_c$ 
& $\rho$   & $-\frac{3}{2}$ & $600$ \\
& $\omega$ & $\frac{1}{2}$ & $600$ \\
& $\sigma$ & $2$ & $600$ \\
$\bar{D}^*\Xi_c\to\bar{D}\Xi_c^*$ 
& $\pi$    & $-\frac{3}{4}\sqrt{2}$ & $600$ \\
& $\eta$   & $\frac{1}{4}\sqrt{2}$ & $600$ \\
& $\rho$   & $-\frac{3}{4}\sqrt{2}$ & $600$ \\
& $\omega$ & $\frac{1}{4}\sqrt{2}$ & $600$ \\
$\bar{D}^*\Xi_c\to\bar{D}^*\Xi_c'$ 
& $\pi$    & $-\frac{3}{4}\sqrt{2}$ & $600$ \\
& $\eta$   & $\frac{1}{4}\sqrt{2}$ & $600$ \\
& $\rho$   & $-\frac{3}{4}\sqrt{2}$ & $600$ \\
& $\omega$ & $\frac{1}{4}\sqrt{2}$ & $600$ \\
$\bar{D}^*\Xi_c\to\bar{D}^*\Xi_c^*$ 
& $\pi$    & $-\frac{3}{4}\sqrt{2}$ & $600$ \\
& $\eta$   & $\frac{1}{4}\sqrt{2}$ & $600$ \\
& $\rho$   & $-\frac{3}{4}\sqrt{2}$ & $600$ \\
& $\omega$ & $\frac{1}{4}\sqrt{2}$ & $600$ \\
$\bar{D}\Xi_c^*\to\bar{D}\Xi_c^*$ 
& $\rho$   & $-\frac{3}{4}$ & $700$ \\
& $\omega$ & $\frac{1}{4}$ & $700$ \\
& $\sigma$ & $1$ & $700$ \\
$\bar{D}\Xi_c^*\to\bar{D}^*\Xi_c'$ 
& $\pi$    & $-\frac{3}{4}$ & $700$ \\
& $\eta$   & $-\frac{1}{12}$ & $700$ \\
& $\rho$   & $-\frac{3}{4}$ & $700$ \\
& $\omega$ & $\frac{1}{4}$ & $700$ \\
$\bar{D}\Xi_c^*\to\bar{D}^*\Xi_c^*$ 
& $\pi$    & $-\frac{3}{4}$ & $700$ \\
& $\eta$   & $-\frac{1}{12}$ & $700$ \\
& $\rho$   & $-\frac{3}{4}$ & $700$ \\
& $\omega$ & $\frac{1}{4}$ & $700$ \\
$\bar{D}^*\Xi_c'\to\bar{D}^*\Xi_c'$ 
& $\pi$    & $-\frac{3}{4}$ & $700$ \\
& $\eta$   & $-\frac{1}{12}$ & $700$ \\
& $\rho$   & $-\frac{3}{4}$ & $700$ \\
& $\omega$ & $\frac{1}{4}$ & $700$ \\
& $\sigma$ & $1$ & $700$ \\
$\bar{D}^*\Xi_c'\to\bar{D}^*\Xi_c^*$ 
& $\pi$    & $-\frac{3}{4}$ & $700$ \\
& $\eta$   & $-\frac{1}{12}$ & $700$ \\
& $\rho$   & $-\frac{3}{4}$ & $700$ \\
& $\omega$ & $\frac{1}{4}$ & $700$ \\
$\bar{D}^*\Xi_c^*\to\bar{D}^*\Xi_c^*$ 
& $\pi$    & $-\frac{3}{4}$ & $700$ \\
& $\eta$   & $-\frac{1}{12}$ & $700$ \\
& $\rho$   & $-\frac{3}{4}$ & $700$ \\
& $\omega$ & $\frac{1}{4}$ & $700$ \\
& $\sigma$ & $1$ & $700$ \\
  \end{tabular}
    \end{ruledtabular}
\end{table}

The Feynman amplitude for a one-meson exchange diagram can be
expressed as   
\begin{align}
  \mathcal{A}_{\lambda'_1\lambda'_2,\lambda_1\lambda_2}
  = \mathrm{IS} \,F^2(q^2)\,\Gamma_{\lambda'_1\lambda'_2}(p'_1,p'_2)
  \mathcal P(q)\Gamma_{\lambda_1\lambda_2}(p_1,p_2) ,
\end{align}
where $\lambda_i$ and $p_i$ represent the helicity and momentum of the
corresponding particle, and $q$ is the momentum of the exchanged
particle. The IS factor corresponds to the SU(3) Clebsch-Gordan
coefficient and isospin factor, and is tabulated in Table~\ref{tab:1}
for each exchange diagram. The vertex $\Gamma$ is obtained from the
effective Lagrangian previously described, and the propagators for the
spin-0 and spin-1 mesons are given by 
\begin{align}
  \mathcal{P}(q) &= \frac{1}{q^2-m^2},\;\;\;
  \mathcal{P}_{\mu\nu}(q) = \frac{1}{q^2-m^2}
  \left(-g_{\mu\nu}+\frac{q_\mu q_\nu}{m^2}\right).
\end{align}

For simplicity, we employ the static propagator for pion exchange,
given by $\mathcal{P}_\pi(q) = -1/(\bm{q}^2 + m_\pi^2)$. For the
heavy-meson propagators, we adopt the same form as for light mesons,
since the heavy-quark mass is finite. Furthermore, parity invariance
reduces the number of contributing processes. The parity relation is
expressed as  
\begin{align}
  \mathcal{A}_{-\lambda'_1-\lambda'_2,-\lambda_1-\lambda_2} =
  \eta(\eta')^{-1}\,
  \mathcal{A}_{\lambda'_1\lambda'_2,\lambda_1\lambda_2},  
    \label{eq:ampi}
\end{align}
where $\eta$ $\eta'$ are defined as 
\begin{align}
    \eta = \eta_1\eta_2(-1)^{J-s_1-s_2}, \hspace{0.5 cm} \eta' =
  \eta_1'\eta_2'(-1)^{J-s_1'-s_2'}. 
\end{align}
Here, $\eta_i$ and $s_i$ designate respectively the intrinsic parity
and spin of the particle, and $J$ denotes the total angular momentum.    

To account for the finite size of hadrons, we introduce a form factor
at each vertex. We adopt the following
parametrization~\cite{Kim:1994ce}  
\begin{align}
  F(q^2) = \left(\frac{n\Lambda^2-m^2}
  {n\Lambda^2-q^2}
  \right)^n,
\label{eq:13}
\end{align}
where $n$ is determined by the momentum power in the vertex. 
This parametrization has the advantage that adjusting $\Lambda$ is not
necessary when varying $n$. It is noteworthy that, in the limit $n \to
\infty$, Eq.~\eqref{eq:13} reduces to a Gaussian form. Although the
cutoff masses $\Lambda$ in Eq.~\eqref{eq:13} are not experimentally
determined for heavy hadron processes, we implement a strategy to
minimize the associated uncertainties. Recent studies have shown that
heavy hadrons possess more compact structures than their light
counterparts~\cite{Kim:2018nqf, Kim:2021xpp}, suggesting that the
cutoff masses for heavy hadrons should be larger than those for light
hadrons. Accordingly, we define the cutoff mass as $\Lambda =
\Lambda_0 + m$, where $m$ denotes the mass of the exchanged meson. We
choose $\Lambda_0$ values in the range of approximately $500$–$700$
MeV for each channel, as summarized in Table~\ref{tab:1}, allowing for
a minimal fitting procedure. 

To simplify the numerical calculations and clarify the spin-parity
assignments for $P_{c\bar{c}s}$ states, we perform a partial-wave
expansion of the $\mathcal{V}$ and $\mathcal{T}$ matrices. This
results in a one-dimensional integral equation: 
\begin{align}
  \mathcal{T}^{J(fi)}_{\lambda'\lambda} (\mathrm{p}',\mathrm{p}) = 
  \mathcal{V}^{J(fi)}_{
  \lambda'\lambda} (\mathrm{p}',\mathrm{p}) + \frac{1}{(2\pi)^3}
  \sum_{k,\lambda_k}\int
  \frac{\mathrm{q}^2d\mathrm{q}}{2E_{k1}E_{k2}}
  \mathcal{V}^{J(fk)}_{\lambda'\lambda_k}(\mathrm{p}',
  \mathrm{q})\frac{E_k}{
  s-E_k^2+i\varepsilon} \mathcal{T}^{J(ki)}_{\lambda_k\lambda}
  (\mathrm{q},\mathrm{p}),
  \label{eq:BS-1d}
\end{align}
where $\lambda'=\{\lambda'_1,\lambda'_2\}$,
$\lambda=\{\lambda_1,\lambda_2\}$ and
$\lambda_k=\{\lambda_{k1},\lambda_{k2}\}$ denote the helicities of the
final ($f$), initial ($i$) and intermediate ($k$) states. The
variables $\mathrm{p}'$, $\mathrm{p}$ and $\mathrm{q}$ represent the
magnitudes of momentum vectors $\bm{p}'$, $\bm{p}$ and $\bm{q}$,
respectively. The partial-wave kernel amplitudes
$\mathcal{V}_{\lambda'\lambda}^{J(fi)}$ are formulated as  
\begin{equation}
  \mathcal{V}^{J(fi)}_{\lambda'\lambda}(\mathrm{p}',\mathrm{p}) = 
  2\pi \int d( \cos\theta) \,
        d^{J}_{\lambda_1-\lambda_2,\lambda'_1-\lambda'_2}(\theta)\,
        \mathcal{V}^{fi}_{\lambda'\lambda}(\mathrm{p}',\mathrm{p},\theta),
  \label{eq:pwd}
\end{equation} 
where $\theta$ is the scattering angle and
$d^{J}_{\lambda_f\lambda_i}(\theta)$ represents the reduced Wigner $D$
functions.   

The integral equation in Eq.~\eqref{eq:BS-1d} contains a singularity
arising from the two-body propagator $\mathcal{G}$. To handle this
singularity, we isolate and treat its singular component
separately. The regularized integral equation is written as  
\begin{align}
  \mathcal{T}^{fi}_{\lambda'\lambda} (\mathrm{p}',\mathrm{p}) = 
  \mathcal{V}^{fi}_{
  \lambda'\lambda} (\mathrm{p}',\mathrm{p}) + \frac{1}{(2\pi)^3}
  \sum_{k,\lambda_k}\left[\int_0^{\infty}d\mathrm{q}
  \frac{\mathrm{q}E_k}{E_{k1}E_{k2}}\frac{\mathcal{F}(\mathrm{q})
  -\mathcal{F}(\tilde{\mathrm{q}}_k)}{s-E_k^2}+ \frac{1}{2\sqrt{s}}
  \left(\ln\left|\frac{\sqrt{s}-E_k^{\mathrm{thr}}}{\sqrt{s}
  +E_k^{\mathrm{thr}}}\right|-i\pi\right)\mathcal{F}
  (\tilde{\mathrm{q}}_k)\right],
  \label{eq:BS-1d-reg}
\end{align}
with 
\begin{align}
  \mathcal{F}(\mathrm{q})=\frac{1}{2}\mathrm{q}\,
  \mathcal{V}^{fk}_{\lambda'\lambda_k}(\mathrm{p}',
  \mathrm{q})\mathcal{T}^{ki}_{\lambda_k\lambda}(\mathrm{q},\mathrm{p}) ,
\end{align}
and $\tilde{\mathrm{q}}_k$ denotes the momentum $\mathrm{q}$ when
$E_{k1}+E_{k2}=\sqrt{s}$. Regularization is applied only when the
total energy $\sqrt{s}$ exceeds the threshold energy of the $k$-th
channel $E_k^{\mathrm{thr}}$. It is worth noting that the form factors
in the kernel amplitudes $\mathcal{V}$ ensure the unitarity of the
transition amplitudes in the high-momentum region.

To compute the $\mathcal{T}$ matrix numerically in
Eq.~\eqref{eq:BS-1d-reg}, we expand the $\mathcal{V}$ matrix in 
the helicity basis and express it in momentum space, where the momenta
are determined using the Gaussian quadrature method. The $\mathcal{T}$
matrix is then obtained by applying the Haftel–Tabakin matrix
inversion method~\cite{Haftel:1970zz} 
\begin{align}
  \mathcal{T} = \left(1-\mathcal{V}\tilde{\mathcal{G}}\right)^{-1} 
  \mathcal{V}.
\end{align}

The resulting $\mathcal{T}$ matrix is expressed in the helicity basis
and does not possess definite parity. To analyze the parity
assignments of the $P_{c\bar{c}s}$ states, we decompose the transition
amplitudes into partial-wave amplitudes with definite parity: 
\begin{align}
  \mathcal{T}^{J\pm}_{\lambda'\lambda} =
  \frac{1}{2}\left[\mathcal{T}^{J}_{\lambda'\lambda} \pm
  \eta_1\eta_2(-1)^{s_1+s_2+\frac{1}{2}}
  \mathcal{T}^{J}_{\lambda'-\lambda}\right], 
\end{align}
where $\mathcal{T}^{J\pm}$ denotes the partial-wave transition
amplitude with total angular momentum $J$ and parity $(-1)^{J\pm
  1/2}$. The factor $1/2$ ensures that no additional factor is needed
when transforming back to the partial-wave component:  
\begin{align}
    \mathcal{T}^{J}_{\lambda'\lambda} =
  \mathcal{T}^{J+}_{\lambda'\lambda} +\mathcal{T}^{J-}_{\lambda'\lambda}.
\end{align}

We emphasize that it is unnecessary to decompose the partial-wave
component with definite parity in Eq.~\eqref{eq:BS-1d}, as parity
invariance is already incorporated in both the effective Lagrangian
and the amplitude calculations, as shown in Eq.~\eqref{eq:ampi}.  
To investigate the dynamical generation of resonances, we express the
$\mathcal{T}$ matrix in the $IJL$ particle
basis~\cite{Machleidt:1987hj}. The relations between the $\mathcal{T}$
matrix elements in the two bases are given by 
\begin{align}
  \mathcal{T}^{JS'S}_{L'L} = \frac{\sqrt{(2L+1)(2L'+1)}}{2J+1}
  \sum_{\lambda'_1\lambda'_2\lambda_1\lambda_2}
  \left(L'0S'\lambda'|J\lambda'\right)
  \left(s'_1\lambda'_1s'_2-\lambda'_2|S'\lambda'\right)
  \left(L0S\lambda|J\lambda\right)
  \left(s_1\lambda_1s_2-\lambda_2|S\lambda\right)
  \mathcal{T}^{J}_{\lambda'_1\lambda'_2,\lambda_1\lambda_2} .
\end{align}
In this work, we present only the diagonal part $\mathcal{T}^{JS}_{L}$
as it is most relevant to particle production. 

\section{Results and discussions \label{sec:3}}
The singly-strange hidden-charm pentaquark was first observed by the
LHCb Collaboration in the decay $\Xi_b^- \to
J/\psi\,\Lambda\,K^-$~\cite{LHCb:2020jpq}. The Collaboration
identified a hidden-charm pentaquark state with strangeness $S=-1$,
denoted as $P_{c\bar{c}s}(4459)$. However, due to limited data, one could not
rule out the possibility that the observed signal originated from two
distinct states. We find a narrow peak structure around 4.40~GeV, 
accompanied by a slightly broader feature nearby. Based on these
results, it is anticipated that additional $P_{c\bar{c}s}$
candidates will be discovered by measuring the $\Xi_b^- \to
J/\psi\,\Lambda\,K^-$ more precisely. Very recently, the Belle
Collaboration confirmed the existence of
$P_{c\bar{c}s}(4459)$~MeV~\cite{Belle:2025pey} with a slightly larger
mass.  

Two years later, the LHCb Collaboration reported the discovery of
a new $P_{c\bar{c}s}$ state in a different decay mode, specifically
$B^- \to J/\psi\,\Lambda\,\bar{p}$~\cite{LHCb:2022ogu}. It is located
at around 4.34~GeV just below the $\bar{D}\Xi_c$ threshold.  
The spin-paritty of the newly observed state was assigned to be
$J^P=1/2^-$, making it the first hidden-charm pentaquark with an
experimentally established spin-parity. Numerous studies have 
suggested that $P_{c\bar{c}s}(4338)$ is associated with a molecular
$\bar{D}\Xi_c$ state with $J^P = 1/2^-$~\cite{Karliner:2022erb,
  Wang:2022mxy, Zhu:2022wpi, Giachino:2022pws}. Additionally, LHCb
investigated the possibility of a narrow resonance near the
$\bar{D}_s\Lambda_c$ threshold; however, this signal was not found to
be statistically significant. 

In this section, we will show how the $P_{c\bar{c}s}$
states are generated dynamically by using the present formalism. 
We then discuss the nature of these singly strange hidden-charm
pentaquarks. We restrict our discussion to the case of zero
total isospin since we only consider the $P_{c\bar{c}s}$ production in the 
$J/\psi\Lambda$ final state. In Table~\ref{tab:1}, we present the IS
factors and cutoff masses for each exchange in the respective
channels. As in the case of the non-strange hidden-charm pentaquarks,
we employ smaller cutoff mass values for transitions to lower channels
($J/\psi\Lambda$, $\bar{D}_s\Lambda_c$, $\bar{D}\Xi_c$,
$\bar{D}_s^*\Lambda_c$, $\bar{D}\Xi_c'$, and $\bar{D}^*\Xi_c$). This
adjustment is necessary to reproduce the experimentally observed
hidden-charm pentaquark states $P_{c\bar{c}s}(4338)$ and
$P_{c\bar{c}s}(4459)$ with $S=-1$. 

\subsection{Negative parity states}
\begin{figure}[htbp]
  \centering
  \includegraphics[scale=0.52]{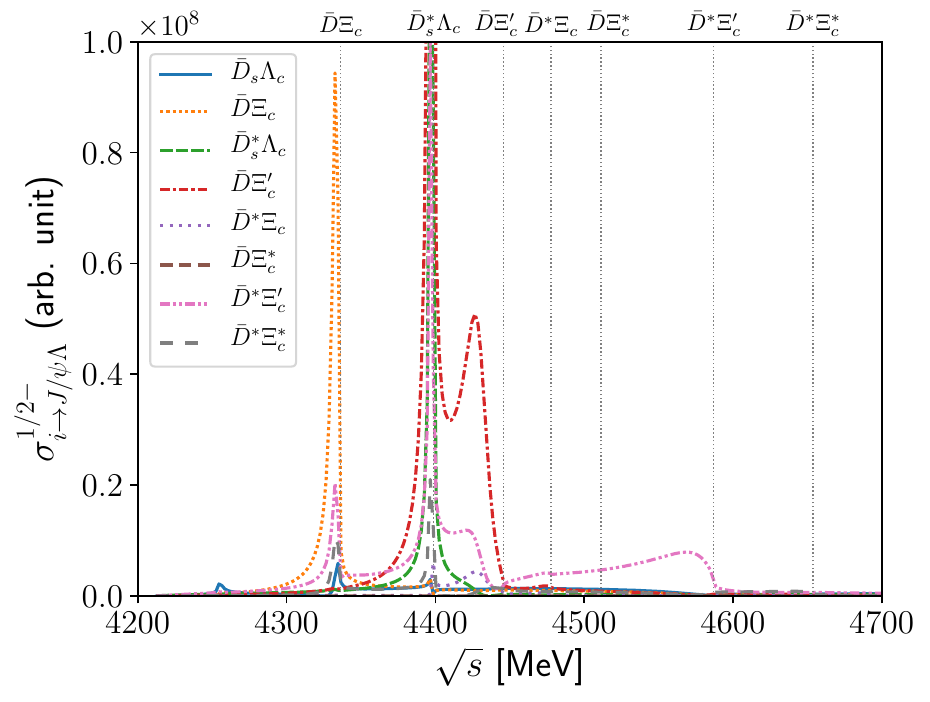}
  \includegraphics[scale=0.52]{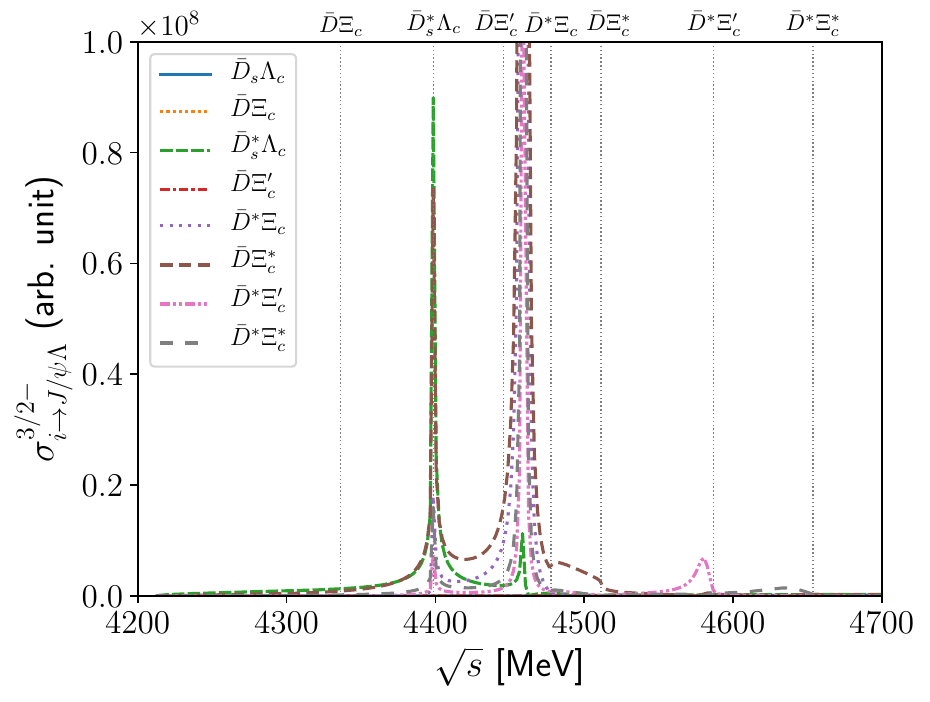}
  \includegraphics[scale=0.52]{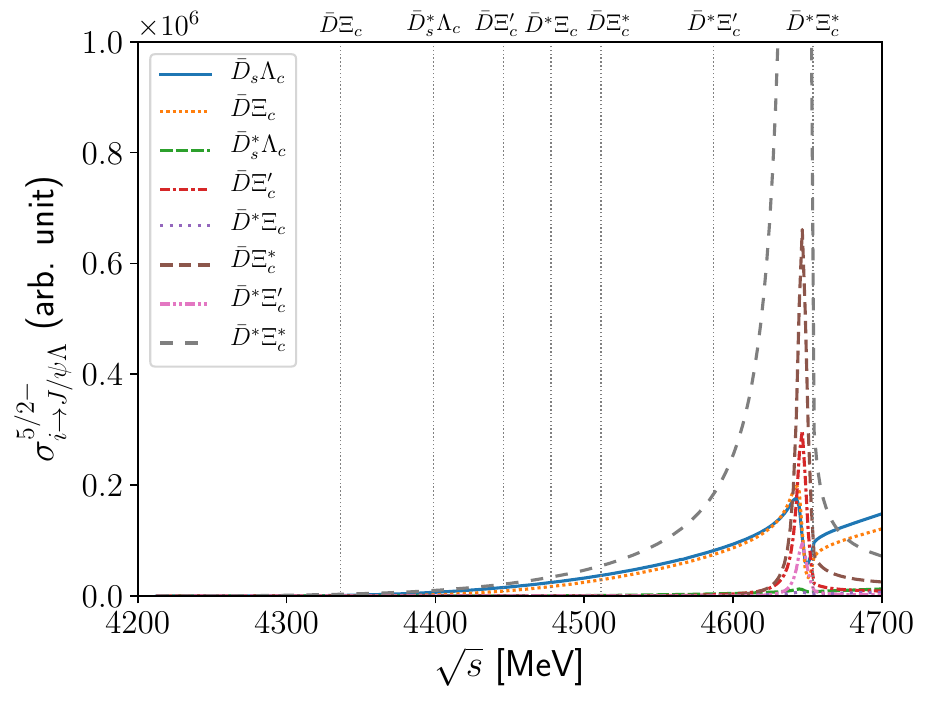}
  \caption{The partial-wave cross sections for the given total
    angular momenta $J=1/2,3/2,5/2$ with negative parity, which
    correspond to the spins and parities of $P_{c\bar{c}s}$, as
    functions of the total energy.}   
  \label{fig:4} 
\end{figure}
We begin by discussing the numerical results for the $P_{c\bar{c}s}$
states with negative parity. As previously mentioned, while
$P_{c\bar{c}s}(4338)$ most likely has $J^P=1/2^-$, the spin-parity
assignment of $P_{c\bar{c}s}(4459)$ remains uncertain. In the present
work, the results suggest that $P_{c\bar{c}s}(4459)$ has
negative parity as well. 

Figure~\ref{fig:4} displays the partial-wave
cross sections for the transitions of various initial states with
$S=-1$ to the $J/\psi\Lambda$ final state as functions of the
$J/\psi\Lambda$ invariant mass.  
In the $J^P=1/2^-$ channel shown in the upper left panel of
Fig.~\ref{fig:4}, we identify four peaks: a narrow peak below the
$\bar{D}\Xi_c$ threshold, two peaks below the $\bar{D}\Xi_c'$
threshold, and a broad peak below the $\bar{D}^*\Xi_c'$
threshold. While the first peak can be associated with the known
$P_{c\bar{c}s}(4338)$, the remaining peaks have not yet been
discovered experimentally. However, given the large cross-sections of
the two peaks below the $\bar{D}\Xi_c'$ threshold, their experimental
observation seems probable. Examining the LHCb data for the decay of
$\Xi_b^-\to J/\psi\Lambda\bar{p}$~\cite{LHCb:2020jpq}, there are
indications of possible structures around 4.4 GeV in the
$J/\psi\Lambda$ invariant mass spectrum, where two potential peaks
appear--a narrow lower structure and a broader higher one. Although
these features have not been officially reported as resonances by the
LHCb Collaboration, possibly due to limited data or statistical
fluctuations, their presence in the data is noticeable and 
is in agreement with the present theoretical predictions. 
We expect that future experiments can confirm these structures.  

In the $J^P = 3/2^-$ case, depicted in the upper right panel of
Fig.~\ref{fig:3}, we observe a narrow peak below the $\bar{D}^*\Xi_c$
threshold, which can be associated with the $P_{c\bar{c}s}(4459)$
state. Unlike the case of $P_{c\bar{c}}(4440)$ and
$P_{c\bar{c}}(4457)$, which have spin-parity assignments of $J^P =
1/2^-$ and $J^P = 3/2^-$, respectively, no corresponding peak
structure appears below the $\bar{D}^*\Xi_c$ threshold in the $J^P =
1/2^-$ channel, as shown in the upper left panel of
Fig.~\ref{fig:3}. This suggests that $P_{c\bar{c}s}(4459)$ is composed
of a single pole structure. Additionally, we find a peak below the
$\bar{D}^*\Xi_c'$ threshold and a broader peak below the
$\bar{D}^*\Xi_c^*$ threshold, although their relatively small
cross sections may hinder experimental identification. A cusp
structure is also visible at the $\bar{D}_s^*\Lambda_c$ threshold. 
In the $J^P = 5/2^-$ case, shown in the lower panel of
Fig.~\ref{fig:3}, we see only a single narrow peak located just below
the $\bar{D}^*\Xi_c^*$ threshold. 

To understand these structures, we analyze the pole positions in the
second Riemann sheet of the complex energy. All poles found in the
negative-parity channel are listed in Table~\ref{tab:2}. We identify
eight poles, two of which correspond to the experimentally observed
$P_{c\bar{c}s}$ states. The present value for the
$P_{c\bar{c}s}(4338)$ mass slightly underestimates the experimental
one. For the $P_{c\bar{c}s}(4459)$, the calculation yields a considerably
smaller width compared to the experimental result. However, given the
present experimental uncertainties, it remains plausible that the
actual width of the $P_{c\bar{c}s}(4459)$ is narrower than currently
reported. In addition to the known states, we predict six new
negative-parity $P_{c\bar{c}s}$ resonances, as well as several
non-resonant structures, which may be confirmed by future experimental
investigations.

\begin{table}[htbp]
  \caption{\label{tab:2} Pole positions of the hidden-charm pentaquark
    states with $S=-1$.}
  \begin{ruledtabular}
  \centering\begin{tabular}{lcccccr}
   \multirow{2}{*}{$J^P$} & \multirow{2}{*}{Resonance states} & 
   \multicolumn{2}{c}{$\sqrt{s_R}=(M-i\Gamma/2)$ MeV} &
   \multicolumn{3}{c}{Known states} \\ 
   & & $M$ & $\Gamma$ & Name & $M$ & $\Gamma$
   \\\hline
     $1/2^{-}$ 
     & $[\bar{D}\Xi_c]_{S=1/2}$ & $4333.6$ & $4.6$ & $P_{c\bar{c}s}$(4338) & $4338.2\pm 0.8$ & $7.0\pm 1.8$ \\
     & $[\bar{D}\Xi_c']_{S=1/2}$ & $4397.8$ & $0.08$ & $-$ & $-$ & $-$ \\
     & $[\bar{D}\Xi_c']_{S=1/2}$ & $4429.8$ & $19.9$ & $-$ & $-$ & $-$ \\
     & $[\bar{D}^*\Xi_c]_{S=1/2}$ & $4474.9$ & $27.9$ &
      $P_{c\bar{c}s}(4472)$  & $4471.7\pm 4.8$ & $21.9 \pm 13.1$ \\
     $3/2^{-}$ 
     & $[\bar{D}^*\Xi_c]_{S=3/2}$ & $4459.3$ & $2.0$ & $P_{c\bar{c}s}$(4459) & $4458.8^{+5.5}_{-3.1}$ & $17^{+10}_{-9}$ \\
     & $[\bar{D}^*\Xi_c']_{S=3/2}$ & $4581.5$ & $10.2$ & $-$ & $-$ & $-$ \\
     & $[\bar{D}^*\Xi_c^*]_{S=3/2}$ & $4643.1$ & $39.6$ & $-$ & $-$ & $-$ \\
     $5/2^{-}$ 
     & $[\bar{D}^*\Xi_c^*]_{S=5/2}$ & $4646.4$ & $7.6$ & $-$ & $-$ & $-$ \\
    \end{tabular}
  \end{ruledtabular}
\end{table}

In the following subsection, we provide a detailed analysis of both
the known and predicted resonances, as well as the non-resonant
structures, by examining the nature of each resonance state. To
facilitate this analysis, we calculate the coupling strength of each
resonance to the partial-wave components of the relevant two-body
states. The coupling strength is extracted from the residue of the
transition amplitude, which is expressed as 
\begin{align}
  \mathcal{R}_{ab} =\lim_{s\to s_R} (s-s_R)\,\mathcal{T}_{ab}/4\pi = g_ag_b.
  \label{eq:residue}
\end{align}
We divide the residue by a factor of $4\pi$, since we use a different
normalization for the partial-wave decomposition in Eq.~\eqref{eq:pwd}
from the conventional one based on the Legendre
polynomials. Note that the definition of the coupling strength in
Eq.~\eqref{eq:residue} does not allow us to determine its absolute
sign. To establish the relative signs, thus, we take the real
part of the coupling to the lowest threshold channel to be positive. 
The coupling strengths of each resonance to all relevant channels are
listed in Table~\ref{tab:3}. 
\begin{table}[htbp]
  \caption{\label{tab:3}Coupling strengths of the eight $P_{c\bar{c}s}$'s
      with $J^P=1/2^-$, $3/2^-$, and $5/2^-$.}  
   \centering\begin{tabular*}{\linewidth}{@{\extracolsep{\fill}} lcccc}
      \toprule
       $J^P$ & \multicolumn{4}{c}{$1/2^-$} \\
       & $P_{c\bar{c}s}(4338)$ & $P_{c\bar{c}s}(4398)$ & $P_{c\bar{c}s}(4430)$ & $P_{c\bar{c}s}(4472)$ \\
       $\sqrt{s_R}$ [MeV] & $4333.6-i2.3$ & $4397.8-i0.04$ & $4429.8-i10.0$ & $4474.9-i14.0$ \\
      \midrule
      $g_{J/\psi \Lambda({}^2S_J)}$  & $0.08-i0.02$ & $0.1+i0.01$ & $0.04+i0.1$ & $0.03-i0.01$           \\
      $g_{J/\psi \Lambda({}^2D_J)}$  & $-$ & $-$ & $-$ & $-$                                             \\
      $g_{J/\psi \Lambda({}^4S_J)}$  & $-$ & $-$ & $-$ & $-$                                             \\
      $g_{J/\psi \Lambda({}^4D_J)}$  & $0.00-i0.00$ & $0.02+i0.00$ & $0.01-i0.02$ & $0.04-i0.04$         \\
      $g_{\bar{D}_s\Lambda_c({}^2S_J)}$ & $-1.18-i2.05$ & $0.14+i0.13$ & $-0.04+i0.21$ & $0.28+i0.20$    \\
      $g_{\bar{D}_s\Lambda_c({}^2D_J)}$ & $-$ & $-$ & $-$ & $-$                                          \\
      $g_{\bar{D}\Xi_c({}^2S_J)}$ & $9.60+i2.41$ & $0.07+i0.18$ & $-0.08+i0.20$ & $0.20+i0.29$           \\
      $g_{\bar{D}\Xi_c({}^2D_J)}$ & $-$ & $-$ & $-$ & $-$                                                \\
      $g_{\bar{D}_s^*\Lambda_c({}^2S_J)}$ & $-0.27+i0.05$ & $-6.24-i0.07$ & $-0.61-i5.98$ & $4.04+i1.13$ \\
      $g_{\bar{D}_s^*\Lambda_c({}^2D_J)}$ & $-$ & $-$ & $-$ & $-$                                        \\
      $g_{\bar{D}_s^*\Lambda_c({}^4S_J)}$ & $-$ & $-$ & $-$ & $-$                                        \\
      $g_{\bar{D}_s^*\Lambda_c({}^4D_J)}$ & $-0.00+i0.00$ & $0.00+i0.00$ & $0.09-i0.01$ & $0.74+i0.15$   \\
      $g_{\bar{D}\Xi_c'({}^2S_J)}$ & $0.24-i0.01$ & $6.89+i0.08$ & $15.99+i5.04$ & $1.06+i5.15$          \\
      $g_{\bar{D}\Xi_c'({}^2D_J)}$ & $-$ & $-$ & $-$ & $-$                                               \\
      $g_{\bar{D}^*\Xi_c({}^2S_J)}$ & $-0.27+i0.02$ & $0.82-i0.03$ & $-13.72+i3.56$ & $-13.57-i7.87$     \\
      $g_{\bar{D}^*\Xi_c({}^2D_J)}$ & $-$ & $-$ & $-$ & $-$                                              \\
      $g_{\bar{D}^*\Xi_c({}^4S_J)}$ & $-$ & $-$ & $-$ & $-$                                              \\
      $g_{\bar{D}^*\Xi_c({}^4D_J)}$ & $-0.01+i0.00$ & $-0.01+i0.00$ & $-0.12+i0.01$ & $-0.01+i0.04$      \\
      $g_{\bar{D}\Xi_c^*({}^4S_J)}$ & $-$ & $-$ & $-$ & $-$                                              \\
      $g_{\bar{D}\Xi_c^*({}^4D_J)}$ & $-0.01+i0.00$ & $0.05+i0.00$ & $-0.07-i0.04$ & $0.01+i0.01$        \\
      $g_{\bar{D}^*\Xi_c'({}^2S_J)}$ & $-5.78+i0.90$ & $-4.84-i0.43$ & $-8.79-i4.61$ & $-5.38-i2.19$     \\
      $g_{\bar{D}^*\Xi_c'({}^2D_J)}$ & $-$ & $-$ & $-$ & $-$                                             \\
      $g_{\bar{D}^*\Xi_c'({}^4S_J)}$ & $-$ & $-$ & $-$ & $-$                                             \\
      $g_{\bar{D}^*\Xi_c'({}^4D_J)}$ & $-0.26+i0.04$ & $0.16+i0.01$ & $-0.24-i0.14$ & $0.10+i0.06$       \\
      $g_{\bar{D}^*\Xi_c^*({}^2S_J)}$ & $-9.66+i1.50$  & $4.44-i0.57$ & $8.95+i3.63$ & $4.32+i0.19$      \\
      $g_{\bar{D}^*\Xi_c^*({}^2D_J)}$ & $-$ & $-$ & $-$ & $-$                                            \\
      $g_{\bar{D}^*\Xi_c^*({}^4S_J)}$ & $-$ & $-$ & $-$ & $-$                                            \\
      $g_{\bar{D}^*\Xi_c^*({}^4D_J)}$ & $0.24-i0.04$ & $0.17-i0.01$ & $-0.31-i0.15$ & $0.11+i0.04$       \\
      $g_{\bar{D}^*\Xi_c^*({}^6S_J)}$ & $-$ & $-$ & $-$ & $-$                                            \\
      $g_{\bar{D}^*\Xi_c^*({}^6D_J)}$ & $0.87-i0.13$ & $0.10-i0.04$ & $0.17+i0.06$ & $0.08-i0.05$        \\
         \bottomrule
     \end{tabular*}
\end{table}
\begin{table}[htbp]
   \addtocounter{table}{-1}
  \caption{Coupling strengths of the eight $P_{c\bar{c}s}$'s
      with $J^P=1/2^-$, $3/2^-$, and $5/2^-$ (continued).}  
   \centering\begin{tabular*}{\linewidth}{@{\extracolsep{\fill}} lcccc}
      \toprule
       $J^P$ & \multicolumn{3}{c}{$3/2^-$} &  \multicolumn{1}{c}{$5/2^-$} \\
       & $P_{c\bar{c}s}(4459)$ & $P_{c\bar{c}s}(4582)$ & $P_{c\bar{c}s}(4643)$ & $P_{c\bar{c}s}(4646)$ \\
       $\sqrt{s_R}$ [MeV] &  $4459.3-i1.0$ & $4581.5-i5.1$ & $4643.1-i19.8$  & $4646.4-i3.8$ \\
      \midrule
      $g_{J/\psi \Lambda({}^2S_J)}$       & $-$ & $-$ & $-$ & $-$ \\
      $g_{J/\psi \Lambda({}^2D_J)}$       & $0.01+i0.00$ & $0.01-i0.00$ & $0.00+i0.02$ & $0.01+i0.00$  \\
      $g_{J/\psi \Lambda({}^4S_J)}$       & $0.06+i0.02$ & $0.02-i0.01$ & $0.01-i0.03$ & $-$  \\
      $g_{J/\psi \Lambda({}^4D_J)}$       & $0.06+i0.01$ & $0.03-i0.02$ & $0.03-i0.06$ & $0.03+i0.00$  \\
      $g_{\bar{D}_s\Lambda_c({}^2S_J)}$   & $-$ & $-$ & $-$ & $-$ \\
      $g_{\bar{D}_s\Lambda_c({}^2D_J)}$   & $0.03+i0.00$ & $-0.24-i0.07$ & $-0.35-i0.09$ & $-0.38-i0.12$  \\
      $g_{\bar{D}\Xi_c({}^2S_J)}$         & $-$ & $-$ & $-$ & $-$ \\
      $g_{\bar{D}\Xi_c({}^2D_J)}$         & $0.03+i0.00$ & $-0.31-i0.10$ & $-0.37-i0.14$ & $-0.54-i0.21$  \\
      $g_{\bar{D}_s^*\Lambda_c({}^2S_J)}$ & $-$ & $-$ & $-$ & $-$ \\
      $g_{\bar{D}_s^*\Lambda_c({}^2D_J)}$ & $0.06-i0.01$ & $0.08+i0.02$ & $-0.52-i0.07$ & $0.12-i0.02$  \\
      $g_{\bar{D}_s^*\Lambda_c({}^4S_J)}$ & $-0.37-i1.62$ & $1.10+i0.96$ & $3.29+i0.80$ & $-$  \\
      $g_{\bar{D}_s^*\Lambda_c({}^4D_J)}$ & $-0.18-i0.53$ & $0.19+i0.28$ & $0.76+i0.14$ & $-0.17-i0.02$  \\
      $g_{\bar{D}\Xi_c'({}^2S_J)}$        & $-$ & $-$ & $-$ & $-$ \\
      $g_{\bar{D}\Xi_c'({}^2D_J)}$        & $0.06-i0.03$ & $1.58+i0.53$ & $-0.07+i0.01$ & $-1.03-i0.27$  \\
      $g_{\bar{D}^*\Xi_c({}^2S_J)}$       & $-$ & $-$ & $-$ & $-$ \\
      $g_{\bar{D}^*\Xi_c({}^2D_J)}$       & $0.02+0.00$ & $0.25-i0.05$ & $-0.65-i0.25$ & $0.11+i0.09$  \\
      $g_{\bar{D}^*\Xi_c({}^4S_J)}$       & $-15.07+i0.76$ & $0.67+i1.08$ & $2.35+i1.09$ & $-$  \\
      $g_{\bar{D}^*\Xi_c({}^4D_J)}$       & $-5.05+i0.26$ & $0.32+i0.24$ & $0.48+i0.16$ & $-0.13-i0.11$  \\
      $g_{\bar{D}\Xi_c^*({}^4S_J)}$       & $-14.75-i2.00$ & $-2.00+i0.11$ & $-3.51+i1.26$ & $-$  \\
      $g_{\bar{D}\Xi_c^*({}^4D_J)}$       & $-4.98-i0.68$ & $-1.97+i0.35$ & $1.06+i1.43$ & $-2.13-i0.35$  \\
      $g_{\bar{D}^*\Xi_c'({}^2S_J)}$      & $-$ & $-$ & $-$ & $-$ \\
      $g_{\bar{D}^*\Xi_c'({}^2D_J)}$      & $-0.06-i0.01$ & $0.00+i0.10$ & $-0.96+i0.58$ & $0.66-i0.07$  \\
      $g_{\bar{D}^*\Xi_c'({}^4S_J)}$      & $5.87+i0.81$ & $-13.37-i3.54$ & $2.34-i0.25$ & $-$  \\
      $g_{\bar{D}^*\Xi_c'({}^4D_J)}$      & $2.03+i0.28$ & $-4.46-i1.20$ & $0.78-i0.39$ & $-0.85+i0.09$  \\
      $g_{\bar{D}^*\Xi_c^*({}^2S_J)}$     & $-$ & $-$ & $-$ & $-$ \\
      $g_{\bar{D}^*\Xi_c^*({}^2D_J)}$     & $0.76+i0.11$ & $0.04-i0.23$ & $1.05+i0.44$ & $1.83+i0.36$  \\
      $g_{\bar{D}^*\Xi_c^*({}^4S_J)}$     & $11.37+i1.57$ & $0.56-i4.00$ & $-19.92-i7.57$ & $-$  \\
      $g_{\bar{D}^*\Xi_c^*({}^4D_J)}$     & $3.60+i0.49$ & $0.19-i1.30$ & $-6.62-2.49i$ & $-0.00-i0.00$  \\
      $g_{\bar{D}^*\Xi_c^*({}^6S_J)}$     & $-$ & $-$ & $-$                            & $14.09+i2.77$ \\
      $g_{\bar{D}^*\Xi_c^*({}^6D_J)}$     & $0.09+i0.01$ & $0.02-i0.01$ & $0.00+i0.03$ & $6.29+i1.25$  \\
         \bottomrule
     \end{tabular*}
\end{table}
It is essential to know the results for the coupling strengths in
Table~\ref{tab:3}, since they will reveal the nature of the 
hidden-charm pentaquarks with $S=-1$. 

\subsubsection{$P_{c\bar{c}s}(4338)$}
The $P_{c\bar{c}s}(4338)$ was first discovered in the $J/\psi\Lambda$
invariant mass spectrum from the decay $B^- \to
J/\psi\,\Lambda\,\bar{p}$. This state was observed with high 
statistical significance, allowing for a definitive determination of
its spin-parity assignment. Located just below the $\bar{D}\Xi_c$
threshold, it has been widely interpreted in various studies as a
molecular state of $\bar{D}\Xi_c$~\cite{Karliner:2022erb,
  Wang:2022mxy, Zhu:2022wpi, Giachino:2022pws}.  

The first resonance listed in Table~\ref{tab:2} is identified with the
$P_{c\bar{c}s}(4338)$. We find that the pole 
corresponding to this state lies approximately 3~MeV below the
$\bar{D}\Xi_c$ threshold. When considering only the single
$\bar{D}\Xi_c$ channel, it appears as the bound state at about 0.2~MeV
below the threshold. We introduce other channels including the 
$\bar{D}\Xi_c$ one, among which the $\bar{D}\Xi_c$ channel and
$\bar{D}^*\Xi_c^*$ channel dominate over all other channels, as shown
in Table~\ref{tab:3}. It is interesting to see that even though the
$\bar{D}^*\Xi_c^*$ channel has the largest threshold energy, it still
influences on the generation of the $P_{c\bar{c}s}(4338)$ state as a
resonance. Consequently, the pole moves to the second
Riemann sheet, such that it arises as the resonance. 
$P_{c\bar{c}s}(4338)$ is strongly coupled to both the
$\bar{D}\Xi_c$ channel and $\bar{D}^*\Xi_c^*$ channel. Note that the
$\bar{D}^* \Xi_c'$ channel has a sizable contribution to the
$P_{c\bar{c}s}(4338)$ state. Though the $P_{c\bar{c}s}(4338)$
pentaquark state observed by the LHCb Collaboration is positioned
slightly above the $\bar{D}\Xi_c$ threshold, the present result
implies that the $P_{c\bar{c}s}(4338)$ can be considered to be a
$\bar{D}\Xi_c$ bound state, based on the present calculation. 

\subsubsection{$P_{c\bar{c}s}(4459)$}
Concerning the $P_{c\bar{c}s}(4459)$, the LHCb and Belle data show
slight discrepancies. The LHCb Collaboration reported its mass as
$(4458.8 \pm 2.9_{-1.1}^{+4.7})$ MeV with a width of $\Gamma = (17.3
\pm 6.5_{-5.7}^{+8.0})$ MeV, while the Belle Collaboration measured
the mass to be $(4471.7 \pm 4.8 \pm 0.6)$ MeV and the width as $\Gamma
= (21.9 \pm 13.1 \pm 2.7)$ MeV. Interestingly, we found two
$P_{c\bar{c}s}$ states that correspond to those reported by the LHCb
and Belle Collaborations, respectively. These states are located below
the $\bar{D}^* \Xi_c$ threshold, as shown in Fig.~\ref{fig:1} and
Table~\ref{tab:3}. Their masses and widths are obtained to be
$M_{P_{c\bar{c}s}(4459)} = 4459.3$ MeV and
$\Gamma_{P_{c\bar{c}s}(4459)} = 2$ MeV, and $M_{P_{c\bar{c}s}(4472)} =
4474.9$ MeV and $\Gamma_{P_{c\bar{c}s}(4472)} = 28$ MeV,
respectively. This indicates that not only the masses but also the
widths are comparable to the experimental data. Based on the
predictions of the present work, we suggest that the $P_{c\bar{c}s}$
state identified by the LHCb Collaboration should be distinguished
from the one reported by the Belle Collaboration. We therefore
conclude that there exist \emph{two} hidden-charm pentaquark states
with $S = -1$ below the $\bar{D}^* \Xi_c$ threshold.

Though these two pentaquark states are positioned very close each
other, their spins are different. While the $P_{c\bar{c}s}(4459)$ has
spin $3/2$, the $P_{c\bar{c}s}(4472)$ has spin $1/2$. Their parities
are negative. It means that these two poles are the separate
resonances without any two-pole structure. As mentioned above, these
two pentaquark states lie below the $\bar{D}^* \Xi_c$ threshold. Thus,
both the resonances may be considered as molecular states consisting
of the $\bar{D}^*$ with $I(J^P)=1/2(1^-)$ and $\Xi_c$ with
$J^P=1/2(1/2)^+$. To understand this nature, we need to examine how
these two pentaquark states arise from the coupled-channel
interactions. 

\begin{figure}[htbp]
  \centering
  \includegraphics[scale=0.56]{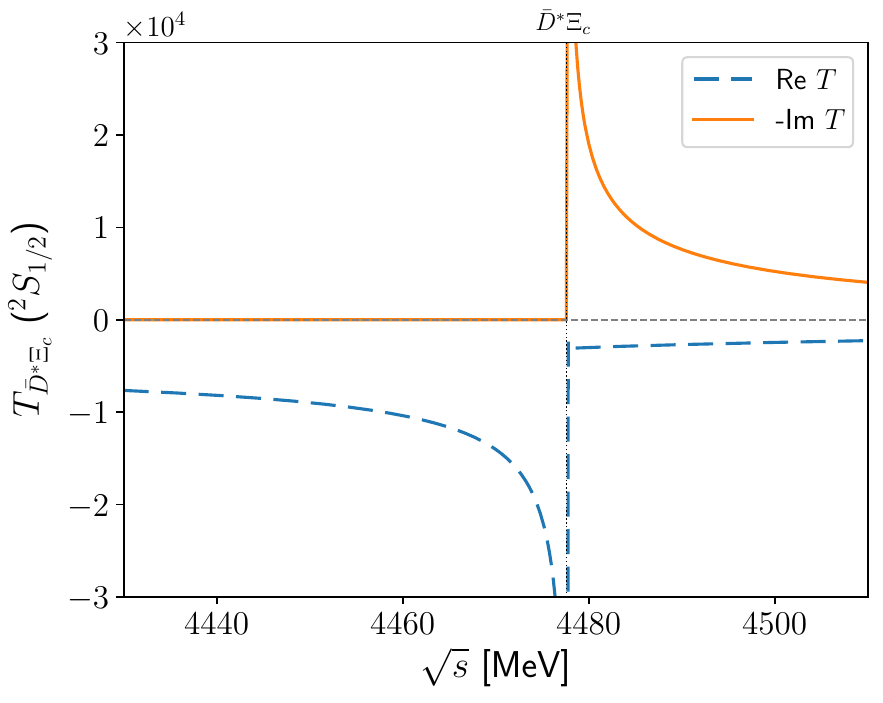}
  \includegraphics[scale=0.56]{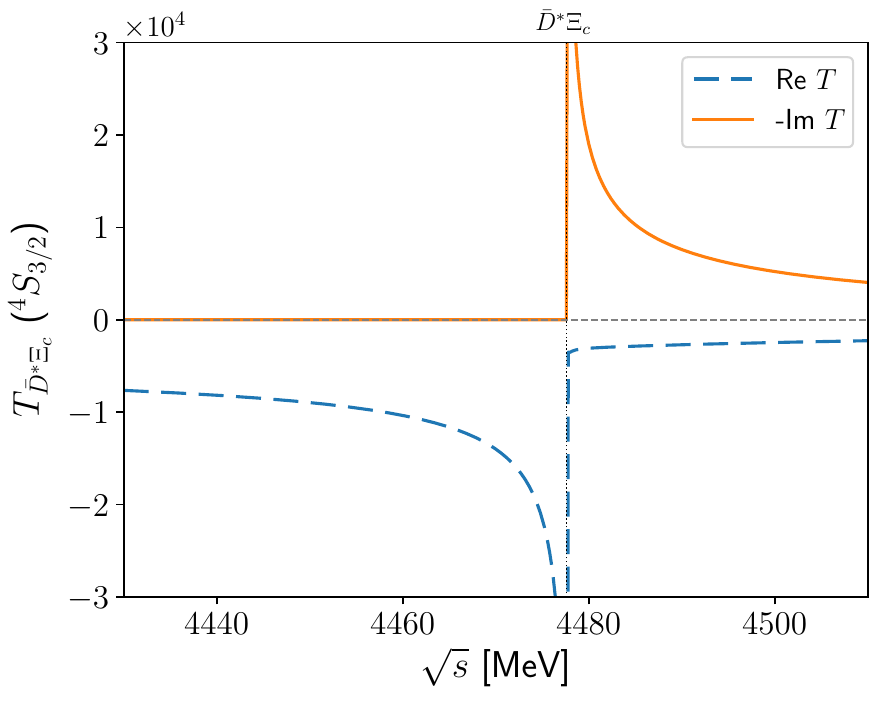}
  \caption{The invariant $\mathcal{T}$ amplitudes for $\bar{D}\Xi_c'$
     elastic scattering with both spin $1/2^-$ and $3/2^-$ as
     functions of the total energy.  generated by the
    single $\bar{D}^*\Xi_c$ channel for 
    $J^P=1/2^-$ (upper panel) and $J^P=3/2^-$ (lower panel).} 
  \label{fig:5}
\end{figure}
As demonstrated in Fig.\ref{fig:5}, the single-channel
$\bar{D}^*\Xi_c$ interaction generates threshold enhancements in both
the $J^P = 1/2^-$ and $3/2^-$ channels. When coupled to other
channels, these enhancements evolve into the broad
$P_{c\bar{c}s}(4475)$ and the narrow $P_{c\bar{c}s}(4459)$ states. As
shown in Table\ref{tab:3}, the broad $P_{c\bar{c}s}(4475)$ couples
most strongly to the $\bar{D}^*\Xi_c$ channel, with the magnitude of
the corresponding coupling strength approximately ten times larger
than its next strongest coupling. In contrast, the narrow
$P_{c\bar{c}s}(4459)$ state, while also coupling most strongly to the
$\bar{D}^*\Xi_c$ channel, exhibits a coupling strength comparable to
that to the $\bar{D}\Xi_c^*$ channel. We find that this crucial
interplay between the coupling strengths causes an interesting
feature: the higher-mass pole $P_{c\bar{c}s}(4472)$ is overshadowed by
the presence of the narrower state, $P_{c\bar{c}s}(4459)$. Comparing
the left panel of Fig.~\ref{fig:4} with its right panel, one can see 
this feature. Although the patterns of the coupling strengths for the 
$P_{c\bar{c}s}(4459)$ and $P_{c\bar{c}s}(4472)$ turn out to be
different, the dominant role of the $\bar{D}^*\Xi_c$ channel in both
cases implies that both the $P_{c\bar{c}s}(4459)$ and
$P_{c\bar{c}s}(4475)$ can be regarded as $\bar{D}^*\Xi_c$ molecular
states. 

\subsubsection{Two resonances in the $\bar{D}\Xi_c'$  channel}
Notably, the $P_{c\bar{c}s}(4398)$ peak almost overlaps with the
$\bar{D}_s^* \Lambda_c$ threshold ($E_{\mathrm{th}} \approx 4398$
MeV). The width of the $P_{c\bar{c}s}(4398)$ is extremely narrow -- less
than 1 MeV -- placing it very close to the real energy axis. This
suggests that the $P_{c\bar{c}s}(4398)$ may be an almost bound state
of $\bar{D}_s^*$ and $\Lambda_c$. On the other hand, the
$P_{c\bar{c}s}(4430)$ lies between the $\bar{D}_s^* \Lambda_c$ and
$\bar{D}\Xi_c'$ thresholds, as shown in Fig.\ref{fig:1}. Therefore, to
determine whether the $P_{c\bar{c}s}(4430)$ is a molecular state of
$\bar{D}$ and $\Xi_c'$, it is necessary to examine its coupling
strengths to various channels in detail. 

\begin{figure}[htbp]
   \centering
   \includegraphics[scale=0.63]{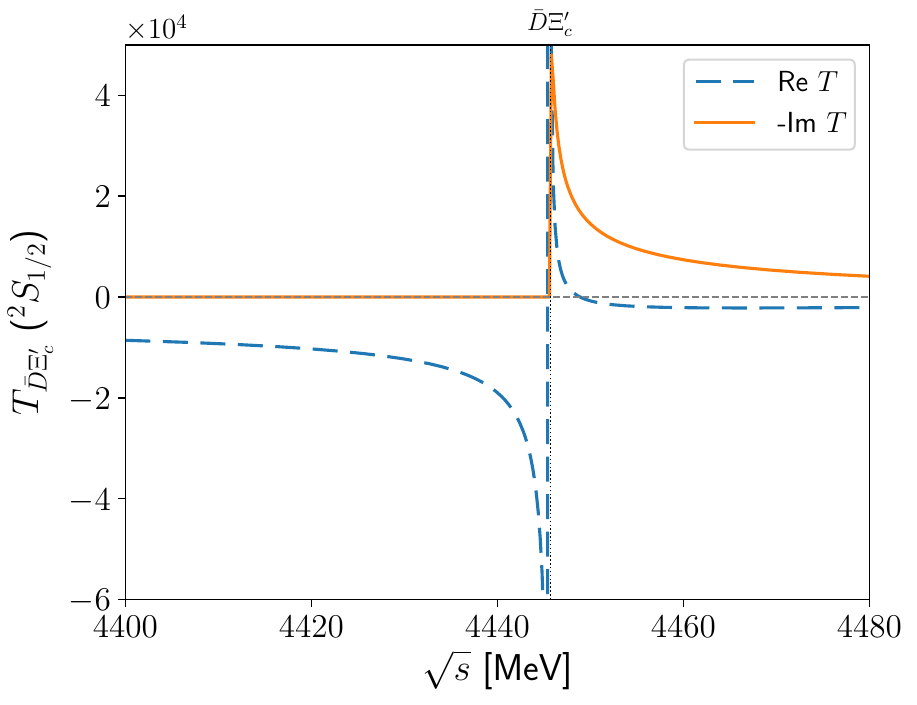}
   \caption{The invariant $\mathcal{T}$ amplitude for $\bar{D}\Xi_c'$
     elastic scattering with $J^P=1/2^-$ as a function of the total
     energy. Note that only the single $\bar{D}\Xi_c'$ channel is
     considered.}   
   \label{fig:6}
\end{figure}
In Fig.\ref{fig:6}, we show that the $\bar{D}\Xi_c'$ single channel
generates a bound state. As additional channels are introduced, this
bound state acquires a finite width and eventually evolves into the
$P_{c\bar{c}s}(4430)$ resonance. As shown in Table\ref{tab:3}, four
different channels contribute to the generation of the
$P_{c\bar{c}s}(4430)$: the $\bar{D}\Xi_c'$, $\bar{D}^*\Xi_c$, $\bar{D}^*
\Xi_c^*$, and $\bar{D}^* \Xi_c'$ channels, among which the
$\bar{D}\Xi_c'$ channel is the most strongly coupled. Therefore, we
may regard the $P_{c\bar{c}s}(4430)$ as a molecular state composed of
a $\bar{D}$ meson and a $\Xi_c'$ baryon. 
   
When all other channels are introduced, the $P_{c\bar{c}s}(4398)$
emerges as a second state, located almost exactly at the
$\bar{D}_s^*\Lambda_c$ threshold. This proximity suggests that it may
be interpreted as a $\bar{D}_s^*\Lambda_c$ bound state. To investigate
its nature, we examine the scattering amplitude generated by the
$\bar{D}_s^*\Lambda_c$ single channel. Interestingly, no bound state is
found below its threshold. This result implies that the
$P_{c\bar{c}s}(4398)$ cannot be interpreted as a pure $\bar{D}\Xi_c'$
or $\bar{D}_s^*\Lambda_c$ molecular state. 

As listed in Table~\ref{tab:3}, four channels are dominantly coupled
to the $P_{c\bar{c}s}(4398)$: the $\bar{D}\Xi_c'$,
$\bar{D}_s^*\Lambda_c$, $\bar{D}^* \Xi_c'$, and $\bar{D}^*\Xi_c^*$
channels. These are the same channels that dominantly contribute to
the $P_{c\bar{c}s}(4430)$ as well. This indicates that the would-be
single pole appearing in the $\bar{D}\Xi_c'$ single channel becomes
split into two hidden-charm pentaquark states with strangeness $S=-1$
due to the interplay of the four aforementioned channels. This
behavior is reminiscent of a possible two-pole structure. 

These two peaks bear a resemblance to the well-known two-pole
structure of the $\Lambda(1405)$\cite{Oller:2000fj,Jido:2003cb}. A
similar structure was also observed in the dynamical generation of the
$b_1(1235)$ axial-vector meson in a previous
work\cite{Clymton:2023txd}. Moreover, the two-pole structure of the
$h_1(1415)$ provides a natural explanation for the conflicting mass
values reported by several experiments for this
state~\cite{Clymton:2024pql}. These observations are not unexpected,
as the two-pole structure is a general feature that arises in the
dynamical generation of resonances via hadron–hadron interactions (see
the recent review~\cite{Meissner:2020khl} for a detailed
discussion). Therefore, it is of great interest to identify this
structure experimentally. 
 
\subsubsection{A resonance in the $\bar{D}^*\Xi_c'$ channel}
In the upper left panel of Fig.\ref{fig:4}, below the $\bar{D}^*\Xi_c'$
threshold, we observe a broad peak in the $\bar{D}^*\Xi_c' \to
J/\psi\Lambda$ transition, which could be misinterpreted as a
resonance state. This is confirmed in Table\ref{tab:2}, where no pole
with $J^P=1/2^-$ is found just below the $\bar{D}^*\Xi_c'$
threshold. Therefore, this broad peak does not correspond to a pole on
the second Riemann sheet. Instead, it originates from a virtual state
near the $\bar{D}^*\Xi_c'$ threshold that affects the physical energy
axis. In contrast, the peak structure in the same transition channel
with $J^P=3/2^-$ corresponds to a genuine resonance, with a mass of 
$4581.5$ MeV and a width of $10.2$ MeV. Although this state has not
yet been experimentally found, it is possible to observe it in the
$\Xi_b^- \to J/\psi\Lambda K^-$ decay channel.

We calculate the scattering amplitude from the 
$\bar{D}^*\Xi_c'$ single channel with $J^P=1/2^-$ and $3/2^-$, and find
that each produces a bound state below its threshold. However, after
coupling to all other channels, only the bound state with $J^P=3/2^-$
arises as a resonance state, while the $J^P=1/2^-$ state becomes
virtual. This phenomenon also appears in our previous work,
specifically where the $\bar{D}^*\Sigma_c^*$ molecular state with
$J^P=1/2^-$ similarly becomes a virtual state due to coupled-channel
effects~\cite{Clymton:2024fbf}. Therefore, it is crucial to consider
all possible coupled channels when constructing the transition
amplitudes. Moreover, these results demonstrate the inadequacy of
considering symmetry alone while neglecting the underlying dynamics. 

\subsubsection{Resonances in the $\bar{D}^*\Xi_c^*$ channel}
The present analysis reveals two additional states that have not yet
been experimentally observed, both appearing near the
$\bar{D}^*\Xi_c^*$ threshold: $P_{c\bar{c}s}(4643)$ and
$P_{c\bar{c}s}(4646)$, with spin-parity assignments of $3/2^-$ and
$5/2^-$, respectively. The lower-mass state exhibits a broader width
than its higher-mass counterpart, and together they generate the
structure observed below the $\bar{D}^*\Xi_c^*$ threshold in the
$\bar{D}^*\Xi_c^*\to J/\psi\,\Lambda$ transition. Notably, no
pole is found below this threshold in the $J^P = 1/2^-$ channel. A 
pattern also is seen previously in the $\bar{D}^*\Xi_c'$ channel 
and in the $\bar{D}^*\Sigma_c^*$ molecular case discussed in our earlier
work~\cite{Clymton:2024fbf}. As shown in the upper right and lower
panels of Fig.~\ref{fig:3}, these resonances produce significantly
smaller cross sections compared to the other states discussed earlier,
making them barely visible. 

To further investigate this phenomenon, we examine the transition 
amplitude constructed from the $\bar{D}^*\Xi_c^*$  single channel for
different total spin states. We identify bound states below the
$\bar{D}^*\Xi_c^*$ threshold with $J^P = 1/2^-$, $3/2^-$, and
$5/2^-$. Upon coupling to all other channels, all of these bound
states evolve into resonances, except the one with $J^P = 1/2^-$. The
$J^P = 3/2^-$ and $5/2^-$ states are identified as the 
$P_{c\bar{c}s}(4643)$ and $P_{c\bar{c}s}(4646)$ resonances,
respectively, as shown in Table~\ref{tab:3}. Both the resonances couple
most strongly to the $\bar{D}^*\Xi_c^*$ channel. The broader width of
the lower-mass state can be attributed to its coupling to a larger
number of channels. Based on these coupling patterns, we conclude that
both resonances are molecular states of $\bar{D}^*\Xi_c^*$. 

\begin{figure}[htbp]
   \centering
   \includegraphics[scale=0.6]{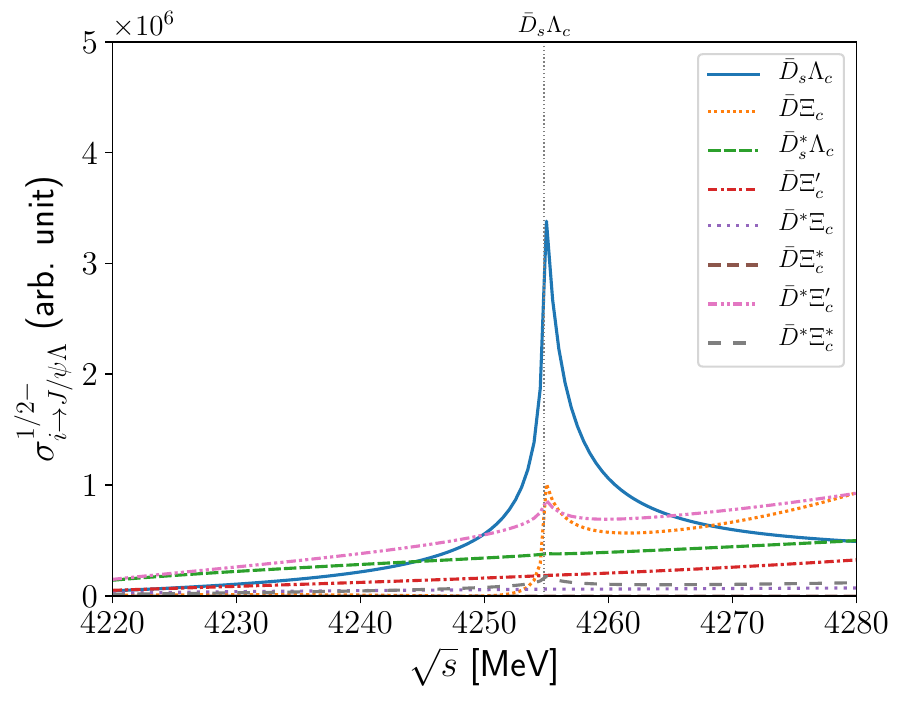}
   \caption{Enlarged view of the upper left panel of
     Fig.~\ref{fig:3} around 4.25 GeV.}    
   \label{fig:7} 
\end{figure}

\subsubsection{Cusps in the $\bar{D}_s\Lambda_c$ and
  $\bar{D}_s^*\Lambda_c$ threshold} 
In addition to resonance and virtual states, the current formalism
also allows for the identification of cusp
structures. Figure~\ref{fig:7}, which presents an enlarged view of the 
upper left panel of Fig.~\ref{fig:4} around 4.25~GeV, reveals a cusp
at the $\bar{D}_s\Lambda_c$ threshold. This cusp is particularly
noteworthy, as it might correspond to the narrow peak observed near the
$\bar{D}_s\Lambda_c$ threshold in the $J/\psi\,\Lambda$ invariant mass
distribution from the $B^- \to J/\psi\,\Lambda\,\bar{p}$ decay
reported in Ref.~\cite{LHCb:2022ogu}, where the interpretation as a
resonance state suffers from lacked statistical significance. The cusp
originates from the $\bar{D}_s\Lambda_c$ single channel, which alone
cannot generate a bound state. The coupled-channel dynamics merely
enhance this threshold effect, leading to a cusp structure rather than
a true resonance. 

Another significant cusp appears in the $J^P = 3/2^-$ channel at the
$\bar{D}_s^*\Lambda_c$ threshold, as shown in the upper right panel of
Fig.~\ref{fig:4}. Although this cusp has an intensity comparable to
that of the $P_{c\bar{c}s}(4338)$ peak, it is obscured by the nearby
$P_{c\bar{c}s}(4398)$ resonance. Its identification would therefore
require a detailed amplitude analysis. 

\subsection{Positive parity}
\begin{figure}[htbp]
   \centering
   \includegraphics[scale=0.5]{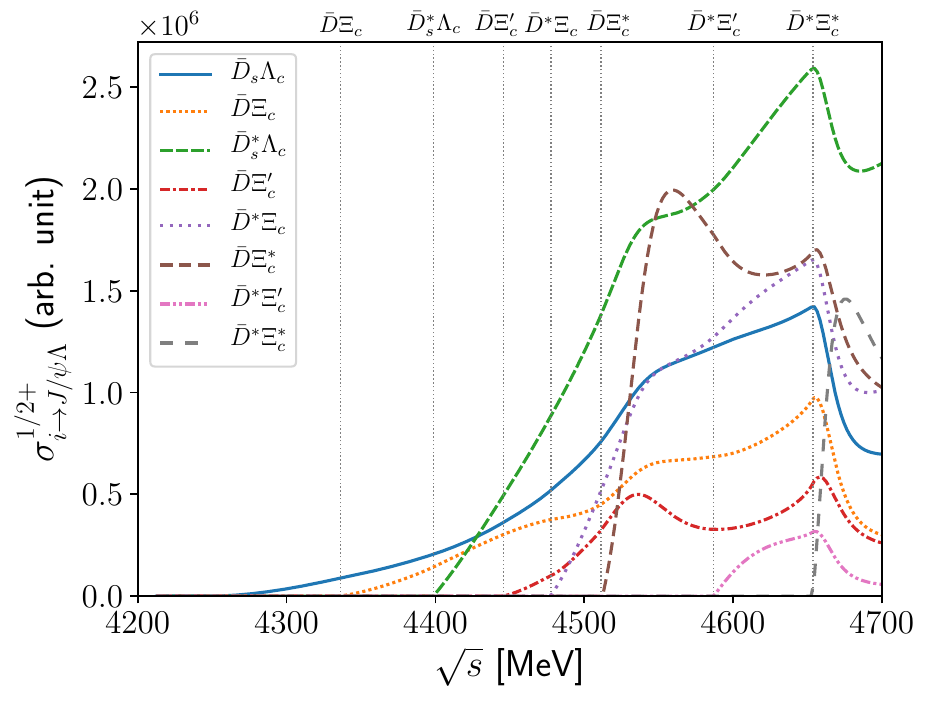}
   \includegraphics[scale=0.5]{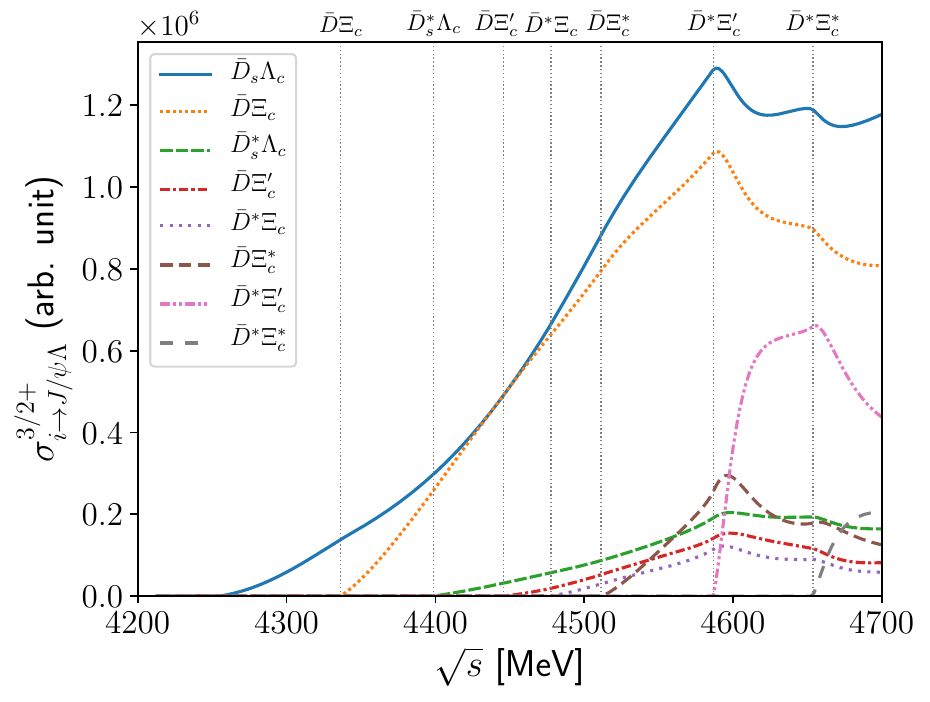}
   \includegraphics[scale=0.5]{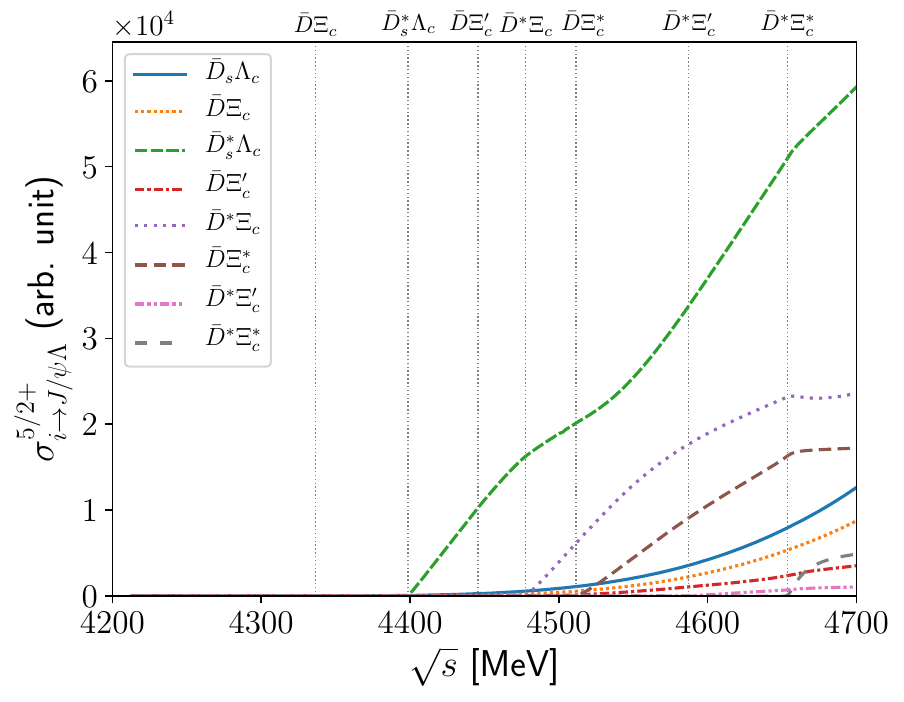}
   \caption{Total energy dependence of the partial-wave total
     cross sections for positive-parity states $(J=1/2, 3/2, 5/2)$
     corresponding to the spin-parity quantum numbers of
     $P_{c\bar{c}s}$.}   
   \label{fig:8}
\end{figure}

The positive-parity states generated in the current work demonstrate
its ability to produce resonances through $P$-wave
interactions. Figure~\ref{fig:8} presents the partial-wave cross
sections for transitions from various initial states to
$J/\psi\,\Lambda$ with $J^P = 1/2^+$, $3/2^+$, and $5/2^+$.  
In the $J^P = 1/2^+$ channel (upper panel), we observe two peak
structures: one located between the $\bar{D}\Xi_c^*$ and
$\bar{D}^*\Xi_c'$ thresholds, which is clearly visible in the
$\bar{D}\Xi_c^* \to J/\psi\,\Lambda$ transition, and another near the
$\bar{D}^*\Xi_c^*$ threshold, which is evident in the $\bar{D}_s^*\Lambda_c\to
J/\psi\,\Lambda$ transition channel.  However, compared to their
negative-parity counterparts, these peaks exhibit smaller cross
sections, indicating that it may be rather difficult to detect them
experimentally. 

The $J^P = 3/2^+$ channel, shown in the upper right panel of
Fig.~\ref{fig:8}, exhibits a distinct peak near the $\bar{D}^*\Xi_c'$
threshold, which is clearly visible in two transition channels:
$\bar{D}_s\Lambda_c \to J/\psi\,\Lambda$ and $\bar{D}\Xi_c\to
J/\psi\,\Lambda$. Additionally, a less prominent bump is observed
around the $\bar{D}^*\Xi_c^*$ threshold. 
In the lower panel of Fig.~\ref{fig:8}, the $J^P = 5/2^+$ channel
shows no significant peak structures. A characteristic feature of
these positive-parity cases is the relatively modest threshold
effects, with most peak structures emerging approximately at their
corresponding thresholds, rather than significantly above or below
them, in contrast to the behavior found in the negative-parity
channels. 

\begin{table}[htbp]
  \caption{\label{tab:4}Coupling strengths of $P_{c\bar{c}s}$'s with
    $J^P=1/2^+$ and $3/2^+$.} 
  \centering\begin{tabular*}{\linewidth}{@{\extracolsep{\fill}} lrrr}
    \toprule
    $J^P$ & \multicolumn{2}{c}{$1/2^+$} & \multicolumn{1}{c}{$3/2^+$} \\
    $\sqrt{s_R}$ [MeV] & $4533.7-i32.4$ & $4658.2-i17.2$ & $4588.9-i20.6$ \\
    \midrule
   $g_{J/\psi\Lambda({}^2P_J)}$            & $0.01+i0.00$ & $0.09+i0.02$ & $0.02-i0.01$ \\
   $g_{J/\psi\Lambda({}^4P_J)}$            & $0.21-i0.10$ & $0.16+i0.03$ & $0.04-i0.00$ \\
   $g_{J/\psi\Lambda({}^4F_J)}$            & $-$ & $-$                    & $0.03-i0.00$ \\
   $g_{\bar{D}_s\Lambda_c({}^2P_J)}$   & $0.02-i0.55$ & $-0.82+i0.34$ & $0.83-i0.44$ \\
   $g_{\bar{D}\Xi_c({}^2P_J)}$         & $0.22-i0.62$ & $-0.96+i0.44$ & $0.86-i0.46$ \\
   $g_{\bar{D}_s^*\Lambda_c({}^2P_J)}$ & $-0.72+i0.52$ & $-0.88+i0.61$ & $0.03-i0.27$ \\
   $g_{\bar{D}_s^*\Lambda_c({}^4P_J)}$ & $-1.02+i1.01$ & $-0.56+i0.42$ & $-0.54+i0.83$ \\
   $g_{\bar{D}_s^*\Lambda_c({}^4F_J)}$ & $-$ & $-$                    & $-0.53+i0.43$ \\
   $g_{\bar{D}\Xi_c'({}^2P_J)}$        & $2.75-i0.35$ & $1.05+i0.69$ & $0.40-i0.07$ \\
   $g_{\bar{D}^*\Xi_c({}^2P_J)}$       & $0.47+i1.07$ & $-1.38+i0.53$ & $-0.13+i0.13$ \\
   $g_{\bar{D}^*\Xi_c({}^4P_J)}$       & $-0.75+i1.90$ & $-0.91+i0.41$ & $-1.27+i0.86$ \\
   $g_{\bar{D}^*\Xi_c({}^4F_J)}$       & $-$ & $-$                    & $-1.69+i0.25$ \\
   $g_{\bar{D}\Xi_c^*({}^4P_J)}$       & $-3.48+i6.22$ & $-0.91-i0.44$ & $-2.81-i0.43$ \\
   $g_{\bar{D}\Xi_c^*({}^4F_J)}$       & $-$ & $-$                    & $-0.02-i1.87$ \\
   $g_{\bar{D}^*\Xi_c'({}^2P_J)}$      & $0.00+i0.00$ & $-1.44+i0.31$ & $-0.43+i4.33$ \\
   $g_{\bar{D}^*\Xi_c'({}^4P_J)}$      & $0.00+i0.00$ & $-0.17+i0.87$ & $5.86-i4.39$ \\
   $g_{\bar{D}^*\Xi_c'({}^4F_J)}$      & $-$ & $-$                    & $3.20-i3.12$ \\
   $g_{\bar{D}^*\Xi_c^*({}^2P_J)}$     & $0.00+i0.00$ & $3.10-i6.75$ & $0.00+i0.00$ \\
   $g_{\bar{D}^*\Xi_c^*({}^4P_J)}$     & $0.00+i0.00$ & $1.99-i2.61$ & $0.00+i0.00$ \\
   $g_{\bar{D}^*\Xi_c^*({}^4F_J)}$     & $-$ & $-$                   & $0.00+i0.00$ \\
   $g_{\bar{D}^*\Xi_c^*({}^6P_J)}$     & $-$ & $-$                   & $0.00+i0.00$ \\
   $g_{\bar{D}^*\Xi_c^*({}^6F_J)}$     & $0.00+i0.01$ & $0.83-i0.51$ & $0.00+i0.00$ \\
   \bottomrule
  \end{tabular*}
\end{table}
Among the three peaks and one bump observed in the $J^P = 1/2^+$ and
$3/2^+$ channels, we identify three corresponding poles, with their
positions and coupling strengths to all channels listed in
Table~\ref{tab:4}. The first pole, $P_{c\bar{c}s}(4534)$, which
generates the initial peak structure in the upper panel of
Fig.~\ref{fig:8}, couples most strongly to the $\bar{D}\Xi_c^*$
channel. Notably, it does not couple to channels with thresholds above
its mass, in stark contrast to the negative-parity cases, where
dynamically generated poles typically exhibit significant couplings to
higher-threshold channels. These characteristics suggest that the
positive-parity hidden-charm pentaquark states may originate from
\emph{genuine} pentaquark states or could merely reflect
coupled-channel effects. 

The second pole, $P_{c\bar{c}s}(4658)$, responsible for the second
peak structure, lies above the highest threshold and couples to all
accessible channels. It exhibits the strongest coupling to the
$\bar{D}^*\Xi_c^*$ channel, particularly in the ${}^2P_{1/2}$ partial
wave. 

A unique feature of the singly-strange hidden-charm pentaquark
system, in contrast to the $P_{c\bar{c}}$ case discussed in our
previous work~\cite{Clymton:2024fbf}, is the presence of a resonance
with $J^P = 3/2^+$. This pole generates the peak structure near the
$\bar{D}^*\Xi_c'$ threshold and couples most strongly to the
${}^4P_{3/2}$ wave of the $\bar{D}^*\Xi_c'$ state, while exhibiting no
coupling to the $\bar{D}^*\Xi_c^*$ channel.

\section{Summary and conclusions\label{sec:4}}
In this work, we have investigated the molecular nature of
singly-strange hidden-charm pentaquark states, $P_{c\bar{c}s}$, using
an off-shell coupled-channel formalism based on effective Lagrangians
that respect heavy-quark spin symmetry, SU(3) flavor symmetry, and
hidden local symmetry. We included all relevant two-body
channels composed of ground-state anti-charmed mesons and
singly-charmed baryons with strangeness $S=-1$, along with the $J/\psi
\Lambda$ channel. 

Solving the coupled-channel scattering equations, we identified eight
negative-parity resonances—four with $J^P=1/2^-$, three with
$J^P=3/2^-$, and one with $J^P=5/2^-$—as well as three positive-parity
states. Among these, the $P_{c\bar{c}s}(4338)$ and
$P_{c\bar{c}s}(4459)$ can be associated with experimentally observed
pentaquark candidates. We have analyzed their strong couplings to
specific meson-baryon channels and interpreted them as hadronic
molecules: the $P_{c\bar{c}s}(4338)$ as a predominantly $\bar{D}\Xi_c$
bound state, and the $P_{c\bar{c}s}(4459)$ as a $\bar{D}^*\Xi_c$
molecular resonance with $J^P=3/2^-$. 

A particularly important result is the identification of the
$P_{c\bar{c}s}(4472)$, located close to the $P_{c\bar{c}s}(4459)$ but
with a larger width and a different spin-parity assignment. Both
originate from the same $\bar{D}^*\Xi_c$ single channel, with
$J^P=3/2^-$ for the $P_{c\bar{c}s}(4459)$ and $J^P=1/2^-$ for the
$P_{c\bar{c}s}(4472)$. This implies that the two observed structures
in the vicinity of 4.46 GeV can be understood as spin-partner states
dynamically generated by the same interaction kernel. The narrow width
and strong coupling of the $P_{c\bar{c}s}(4459)$ to the
$J/\psi\Lambda$ channel are consistent with the recent observation by
LHCb, while the broader $P_{c\bar{c}s}(4472)$ is potentially relevant
for the structure reported by the Belle Collaboration. These findings
strongly support the molecular interpretation of the observed
$P_{c\bar{c}s}$ candidates. 

On the other hand, the two resonances $P_{c\bar{c}s}(4398)$ and
$P_{c\bar{c}s}(4430)$, located near the $\bar{D}_s^*\Lambda_c$ and
$\bar{D}\Xi_c'$ thresholds, respectively, show signatures of a
two-pole behavior. This phenomenon is reminiscent of the well-known
$\Lambda(1405)$ and other mesonic states such as the $b_1(1235)$ and
$h_1(1415)$, where two poles emerge due to channel coupling. The
$P_{c\bar{c}s}(4398)$ lies extremely close to the real axis with a
narrow width, while the $P_{c\bar{c}s}(4430)$ appears as a broader
resonance. Our analysis of the scattering amplitudes and channel
couplings indicates that both states originate from a single-channel
pole in the $\bar{D}\Xi_c'$ system, which splits due to the dynamical
effects of other channels. 

We also analyzed virtual state effects, particularly in the
$J^P=1/2^-$ channel of the $\bar{D}^*\Xi_c'$ system, where no physical
pole was found despite a visible enhancement near threshold. In
contrast, a genuine resonance with $J^P=3/2^-$ and mass $4581.5$ MeV
was identified in the same channel. This resonance is predicted to be
detectable in the $\Xi_b^- \to J/\psi \Lambda K^-$ decay, as its cross
section, while small, remains detectable. 

While we considered the $P_{c\bar{c}s}$ resonances with negative
  parity as molecular states, there is a caveat. To underpin the
  molecular nature of these resonances, one must carefully study their
  decay modes and other physical properties. We will leave it as a
  future work. 
  
We can also extend the current formalism to investigate the $S=-2$
hidden-charm pentaquark states. This will involve the charmed mesons
with strangeness $S=0$ and $-1$, together with the singly-charmed
baryons with $S=-1$ and $-2$, so that we can construct the two-body
meson–baryon scattering amplitudes with $S=-2$. The corresponding work
is under way. 

\begin{acknowledgments}
The present work was supported by the Young Scientist Training (YST)
Program at the Asia Pacific Center for Theoretical Physics (APCTP)
through the Science and Technology Promotion Fund and Lottery Fund of 
the Korean Government and also by the Korean Local Governments –
Gyeongsangbuk-do Province and Pohang City (SC), the Basic Science
Research Program through the National Research Foundation of Korea
funded by the Korean government (Ministry of Education, Science and
Technology, MEST), Grant-No. RS-2025-00513982 (HChK), and the PUTI Q1 
Grant from University of Indonesia under contract
No. NKB-441/UN2.RST/HKP.05.00/2024 (TM).   
\end{acknowledgments}

\bibliography{Pcs}

\begin{thebibliography}{47}%
\makeatletter
\providecommand \@ifxundefined [1]{%
 \@ifx{#1\undefined}
}%
\providecommand \@ifnum [1]{%
 \ifnum #1\expandafter \@firstoftwo
 \else \expandafter \@secondoftwo
 \fi
}%
\providecommand \@ifx [1]{%
 \ifx #1\expandafter \@firstoftwo
 \else \expandafter \@secondoftwo
 \fi
}%
\providecommand \natexlab [1]{#1}%
\providecommand \enquote  [1]{``#1''}%
\providecommand \bibnamefont  [1]{#1}%
\providecommand \bibfnamefont [1]{#1}%
\providecommand \citenamefont [1]{#1}%
\providecommand \href@noop [0]{\@secondoftwo}%
\providecommand \href [0]{\begingroup \@sanitize@url \@href}%
\providecommand \@href[1]{\@@startlink{#1}\@@href}%
\providecommand \@@href[1]{\endgroup#1\@@endlink}%
\providecommand \@sanitize@url [0]{\catcode `\\12\catcode `\$12\catcode
  `\&12\catcode `\#12\catcode `\^12\catcode `\_12\catcode `\%12\relax}%
\providecommand \@@startlink[1]{}%
\providecommand \@@endlink[0]{}%
\providecommand \url  [0]{\begingroup\@sanitize@url \@url }%
\providecommand \@url [1]{\endgroup\@href {#1}{\urlprefix }}%
\providecommand \urlprefix  [0]{URL }%
\providecommand \Eprint [0]{\href }%
\providecommand \doibase [0]{https://doi.org/}%
\providecommand \selectlanguage [0]{\@gobble}%
\providecommand \bibinfo  [0]{\@secondoftwo}%
\providecommand \bibfield  [0]{\@secondoftwo}%
\providecommand \translation [1]{[#1]}%
\providecommand \BibitemOpen [0]{}%
\providecommand \bibitemStop [0]{}%
\providecommand \bibitemNoStop [0]{.\EOS\space}%
\providecommand \EOS [0]{\spacefactor3000\relax}%
\providecommand \BibitemShut  [1]{\csname bibitem#1\endcsname}%
\let\auto@bib@innerbib\@empty
\bibitem [{\citenamefont {Aaij}\ \emph {et~al.}(2015)\citenamefont {Aaij} \emph
  {et~al.}}]{LHCb:2015yax}%
  \BibitemOpen
  \bibfield  {author} {\bibinfo {author} {\bibfnamefont {R.}~\bibnamefont
  {Aaij}} \emph {et~al.} (\bibinfo {collaboration} {LHCb}),\ }\href
  {https://doi.org/10.1103/PhysRevLett.115.072001} {\bibfield  {journal}
  {\bibinfo  {journal} {Phys. Rev. Lett.}\ }\textbf {\bibinfo {volume} {115}},\
  \bibinfo {pages} {072001} (\bibinfo {year} {2015})},\ \Eprint
  {https://arxiv.org/abs/1507.03414} {arXiv:1507.03414 [hep-ex]} \BibitemShut
  {NoStop}%
\bibitem [{\citenamefont {Aaij}\ \emph {et~al.}(2019)\citenamefont {Aaij} \emph
  {et~al.}}]{LHCb:2019kea}%
  \BibitemOpen
  \bibfield  {author} {\bibinfo {author} {\bibfnamefont {R.}~\bibnamefont
  {Aaij}} \emph {et~al.} (\bibinfo {collaboration} {LHCb}),\ }\href
  {https://doi.org/10.1103/PhysRevLett.122.222001} {\bibfield  {journal}
  {\bibinfo  {journal} {Phys. Rev. Lett.}\ }\textbf {\bibinfo {volume} {122}},\
  \bibinfo {pages} {222001} (\bibinfo {year} {2019})},\ \Eprint
  {https://arxiv.org/abs/1904.03947} {arXiv:1904.03947 [hep-ex]} \BibitemShut
  {NoStop}%
\bibitem [{\citenamefont {Aaij}\ \emph {et~al.}(2022)\citenamefont {Aaij} \emph
  {et~al.}}]{LHCb:2021chn}%
  \BibitemOpen
  \bibfield  {author} {\bibinfo {author} {\bibfnamefont {R.}~\bibnamefont
  {Aaij}} \emph {et~al.} (\bibinfo {collaboration} {LHCb}),\ }\href
  {https://doi.org/10.1103/PhysRevLett.128.062001} {\bibfield  {journal}
  {\bibinfo  {journal} {Phys. Rev. Lett.}\ }\textbf {\bibinfo {volume} {128}},\
  \bibinfo {pages} {062001} (\bibinfo {year} {2022})},\ \Eprint
  {https://arxiv.org/abs/2108.04720} {arXiv:2108.04720 [hep-ex]} \BibitemShut
  {NoStop}%
\bibitem [{\citenamefont {Aaij}\ \emph {et~al.}(2023)\citenamefont {Aaij} \emph
  {et~al.}}]{LHCb:2022ogu}%
  \BibitemOpen
  \bibfield  {author} {\bibinfo {author} {\bibfnamefont {R.}~\bibnamefont
  {Aaij}} \emph {et~al.} (\bibinfo {collaboration} {LHCb}),\ }\href
  {https://doi.org/10.1103/PhysRevLett.131.031901} {\bibfield  {journal}
  {\bibinfo  {journal} {Phys. Rev. Lett.}\ }\textbf {\bibinfo {volume} {131}},\
  \bibinfo {pages} {031901} (\bibinfo {year} {2023})},\ \Eprint
  {https://arxiv.org/abs/2210.10346} {arXiv:2210.10346 [hep-ex]} \BibitemShut
  {NoStop}%
\bibitem [{\citenamefont {Adachi}\ \emph {et~al.}(2025)\citenamefont {Adachi}
  \emph {et~al.}}]{Belle:2025pey}%
  \BibitemOpen
  \bibfield  {author} {\bibinfo {author} {\bibfnamefont {I.}~\bibnamefont
  {Adachi}} \emph {et~al.} (\bibinfo {collaboration} {Belle, Belle II}),\
  }\href@noop {} {\  (\bibinfo {year} {2025})},\ \Eprint
  {https://arxiv.org/abs/2502.09951} {arXiv:2502.09951 [hep-ex]} \BibitemShut
  {NoStop}%
\bibitem [{\citenamefont {Hayrapetyan}\ \emph {et~al.}(2024)\citenamefont
  {Hayrapetyan} \emph {et~al.}}]{CMS:2024vnm}%
  \BibitemOpen
  \bibfield  {author} {\bibinfo {author} {\bibfnamefont {A.}~\bibnamefont
  {Hayrapetyan}} \emph {et~al.} (\bibinfo {collaboration} {CMS}),\ }\href
  {https://doi.org/10.1140/epjc/s10052-024-13114-9} {\bibfield  {journal}
  {\bibinfo  {journal} {Eur. Phys. J. C}\ }\textbf {\bibinfo {volume} {84}},\
  \bibinfo {pages} {1062} (\bibinfo {year} {2024})},\ \Eprint
  {https://arxiv.org/abs/2401.16303} {arXiv:2401.16303 [hep-ex]} \BibitemShut
  {NoStop}%
\bibitem [{\citenamefont {Aaij}\ \emph {et~al.}(2025)\citenamefont {Aaij} \emph
  {et~al.}}]{LHCb:2025lhk}%
  \BibitemOpen
  \bibfield  {author} {\bibinfo {author} {\bibfnamefont {R.}~\bibnamefont
  {Aaij}} \emph {et~al.} (\bibinfo {collaboration} {LHCb}),\ }\href@noop {} {\
  (\bibinfo {year} {2025})},\ \Eprint {https://arxiv.org/abs/2501.12779}
  {arXiv:2501.12779 [hep-ex]} \BibitemShut {NoStop}%
\bibitem [{\citenamefont {Wu}\ \emph {et~al.}(2010)\citenamefont {Wu},
  \citenamefont {Molina}, \citenamefont {Oset},\ and\ \citenamefont
  {Zou}}]{Wu:2010jy}%
  \BibitemOpen
  \bibfield  {author} {\bibinfo {author} {\bibfnamefont {J.-J.}\ \bibnamefont
  {Wu}}, \bibinfo {author} {\bibfnamefont {R.}~\bibnamefont {Molina}}, \bibinfo
  {author} {\bibfnamefont {E.}~\bibnamefont {Oset}},\ and\ \bibinfo {author}
  {\bibfnamefont {B.~S.}\ \bibnamefont {Zou}},\ }\href
  {https://doi.org/10.1103/PhysRevLett.105.232001} {\bibfield  {journal}
  {\bibinfo  {journal} {Phys. Rev. Lett.}\ }\textbf {\bibinfo {volume} {105}},\
  \bibinfo {pages} {232001} (\bibinfo {year} {2010})},\ \Eprint
  {https://arxiv.org/abs/1007.0573} {arXiv:1007.0573 [nucl-th]} \BibitemShut
  {NoStop}%
\bibitem [{\citenamefont {Xiao}\ \emph {et~al.}(2019)\citenamefont {Xiao},
  \citenamefont {Nieves},\ and\ \citenamefont {Oset}}]{Xiao:2019gjd}%
  \BibitemOpen
  \bibfield  {author} {\bibinfo {author} {\bibfnamefont {C.~W.}\ \bibnamefont
  {Xiao}}, \bibinfo {author} {\bibfnamefont {J.}~\bibnamefont {Nieves}},\ and\
  \bibinfo {author} {\bibfnamefont {E.}~\bibnamefont {Oset}},\ }\href
  {https://doi.org/10.1016/j.physletb.2019.135051} {\bibfield  {journal}
  {\bibinfo  {journal} {Phys. Lett. B}\ }\textbf {\bibinfo {volume} {799}},\
  \bibinfo {pages} {135051} (\bibinfo {year} {2019})},\ \Eprint
  {https://arxiv.org/abs/1906.09010} {arXiv:1906.09010 [hep-ph]} \BibitemShut
  {NoStop}%
\bibitem [{\citenamefont {Wang}\ \emph {et~al.}(2020)\citenamefont {Wang},
  \citenamefont {Meng},\ and\ \citenamefont {Zhu}}]{Wang:2019nvm}%
  \BibitemOpen
  \bibfield  {author} {\bibinfo {author} {\bibfnamefont {B.}~\bibnamefont
  {Wang}}, \bibinfo {author} {\bibfnamefont {L.}~\bibnamefont {Meng}},\ and\
  \bibinfo {author} {\bibfnamefont {S.-L.}\ \bibnamefont {Zhu}},\ }\href
  {https://doi.org/10.1103/PhysRevD.101.034018} {\bibfield  {journal} {\bibinfo
   {journal} {Phys. Rev. D}\ }\textbf {\bibinfo {volume} {101}},\ \bibinfo
  {pages} {034018} (\bibinfo {year} {2020})},\ \Eprint
  {https://arxiv.org/abs/1912.12592} {arXiv:1912.12592 [hep-ph]} \BibitemShut
  {NoStop}%
\bibitem [{\citenamefont {Du}\ \emph {et~al.}(2021)\citenamefont {Du},
  \citenamefont {Guo},\ and\ \citenamefont {Oller}}]{Du:2021bgb}%
  \BibitemOpen
  \bibfield  {author} {\bibinfo {author} {\bibfnamefont {M.-L.}\ \bibnamefont
  {Du}}, \bibinfo {author} {\bibfnamefont {Z.-H.}\ \bibnamefont {Guo}},\ and\
  \bibinfo {author} {\bibfnamefont {J.~A.}\ \bibnamefont {Oller}},\ }\href
  {https://doi.org/10.1103/PhysRevD.104.114034} {\bibfield  {journal} {\bibinfo
   {journal} {Phys. Rev. D}\ }\textbf {\bibinfo {volume} {104}},\ \bibinfo
  {pages} {114034} (\bibinfo {year} {2021})},\ \Eprint
  {https://arxiv.org/abs/2109.14237} {arXiv:2109.14237 [hep-ph]} \BibitemShut
  {NoStop}%
\bibitem [{\citenamefont {Karliner}\ and\ \citenamefont
  {Rosner}(2022)}]{Karliner:2022erb}%
  \BibitemOpen
  \bibfield  {author} {\bibinfo {author} {\bibfnamefont {M.}~\bibnamefont
  {Karliner}}\ and\ \bibinfo {author} {\bibfnamefont {J.~L.}\ \bibnamefont
  {Rosner}},\ }\href {https://doi.org/10.1103/PhysRevD.106.036024} {\bibfield
  {journal} {\bibinfo  {journal} {Phys. Rev. D}\ }\textbf {\bibinfo {volume}
  {106}},\ \bibinfo {pages} {036024} (\bibinfo {year} {2022})},\ \Eprint
  {https://arxiv.org/abs/2207.07581} {arXiv:2207.07581 [hep-ph]} \BibitemShut
  {NoStop}%
\bibitem [{\citenamefont {Giachino}\ \emph {et~al.}(2023)\citenamefont
  {Giachino}, \citenamefont {Hosaka}, \citenamefont {Santopinto}, \citenamefont
  {Takeuchi}, \citenamefont {Takizawa},\ and\ \citenamefont
  {Yamaguchi}}]{Giachino:2022pws}%
  \BibitemOpen
  \bibfield  {author} {\bibinfo {author} {\bibfnamefont {A.}~\bibnamefont
  {Giachino}}, \bibinfo {author} {\bibfnamefont {A.}~\bibnamefont {Hosaka}},
  \bibinfo {author} {\bibfnamefont {E.}~\bibnamefont {Santopinto}}, \bibinfo
  {author} {\bibfnamefont {S.}~\bibnamefont {Takeuchi}}, \bibinfo {author}
  {\bibfnamefont {M.}~\bibnamefont {Takizawa}},\ and\ \bibinfo {author}
  {\bibfnamefont {Y.}~\bibnamefont {Yamaguchi}},\ }\href
  {https://doi.org/10.1103/PhysRevD.108.074012} {\bibfield  {journal} {\bibinfo
   {journal} {Phys. Rev. D}\ }\textbf {\bibinfo {volume} {108}},\ \bibinfo
  {pages} {074012} (\bibinfo {year} {2023})},\ \Eprint
  {https://arxiv.org/abs/2209.10413} {arXiv:2209.10413 [hep-ph]} \BibitemShut
  {NoStop}%
\bibitem [{\citenamefont {Wang}\ and\ \citenamefont
  {Liu}(2022)}]{Wang:2022mxy}%
  \BibitemOpen
  \bibfield  {author} {\bibinfo {author} {\bibfnamefont {F.-L.}\ \bibnamefont
  {Wang}}\ and\ \bibinfo {author} {\bibfnamefont {X.}~\bibnamefont {Liu}},\
  }\href {https://doi.org/10.1016/j.physletb.2022.137583} {\bibfield  {journal}
  {\bibinfo  {journal} {Phys. Lett. B}\ }\textbf {\bibinfo {volume} {835}},\
  \bibinfo {pages} {137583} (\bibinfo {year} {2022})},\ \Eprint
  {https://arxiv.org/abs/2207.10493} {arXiv:2207.10493 [hep-ph]} \BibitemShut
  {NoStop}%
\bibitem [{\citenamefont {Yan}\ \emph {et~al.}(2023)\citenamefont {Yan},
  \citenamefont {Peng}, \citenamefont {S\'anchez~S\'anchez},\ and\
  \citenamefont {Pavon~Valderrama}}]{Yan:2022wuz}%
  \BibitemOpen
  \bibfield  {author} {\bibinfo {author} {\bibfnamefont {M.-J.}\ \bibnamefont
  {Yan}}, \bibinfo {author} {\bibfnamefont {F.-Z.}\ \bibnamefont {Peng}},
  \bibinfo {author} {\bibfnamefont {M.}~\bibnamefont {S\'anchez~S\'anchez}},\
  and\ \bibinfo {author} {\bibfnamefont {M.}~\bibnamefont {Pavon~Valderrama}},\
  }\href {https://doi.org/10.1103/PhysRevD.107.074025} {\bibfield  {journal}
  {\bibinfo  {journal} {Phys. Rev. D}\ }\textbf {\bibinfo {volume} {107}},\
  \bibinfo {pages} {074025} (\bibinfo {year} {2023})},\ \Eprint
  {https://arxiv.org/abs/2207.11144} {arXiv:2207.11144 [hep-ph]} \BibitemShut
  {NoStop}%
\bibitem [{\citenamefont {Zhu}\ \emph {et~al.}(2023)\citenamefont {Zhu},
  \citenamefont {Kong},\ and\ \citenamefont {He}}]{Zhu:2022wpi}%
  \BibitemOpen
  \bibfield  {author} {\bibinfo {author} {\bibfnamefont {J.-T.}\ \bibnamefont
  {Zhu}}, \bibinfo {author} {\bibfnamefont {S.-Y.}\ \bibnamefont {Kong}},\ and\
  \bibinfo {author} {\bibfnamefont {J.}~\bibnamefont {He}},\ }\href
  {https://doi.org/10.1103/PhysRevD.107.034029} {\bibfield  {journal} {\bibinfo
   {journal} {Phys. Rev. D}\ }\textbf {\bibinfo {volume} {107}},\ \bibinfo
  {pages} {034029} (\bibinfo {year} {2023})},\ \Eprint
  {https://arxiv.org/abs/2211.06232} {arXiv:2211.06232 [hep-ph]} \BibitemShut
  {NoStop}%
\bibitem [{\citenamefont {Shi}\ \emph {et~al.}(2021)\citenamefont {Shi},
  \citenamefont {Huang},\ and\ \citenamefont {Wang}}]{Shi:2021wyt}%
  \BibitemOpen
  \bibfield  {author} {\bibinfo {author} {\bibfnamefont {P.-P.}\ \bibnamefont
  {Shi}}, \bibinfo {author} {\bibfnamefont {F.}~\bibnamefont {Huang}},\ and\
  \bibinfo {author} {\bibfnamefont {W.-L.}\ \bibnamefont {Wang}},\ }\href
  {https://doi.org/10.1140/epja/s10050-021-00542-4} {\bibfield  {journal}
  {\bibinfo  {journal} {Eur. Phys. J. A}\ }\textbf {\bibinfo {volume} {57}},\
  \bibinfo {pages} {237} (\bibinfo {year} {2021})},\ \Eprint
  {https://arxiv.org/abs/2107.08680} {arXiv:2107.08680 [hep-ph]} \BibitemShut
  {NoStop}%
\bibitem [{\citenamefont {Li}\ \emph {et~al.}(2023)\citenamefont {Li},
  \citenamefont {Liu}, \citenamefont {Man}, \citenamefont {Si},\ and\
  \citenamefont {Wu}}]{Li:2023aui}%
  \BibitemOpen
  \bibfield  {author} {\bibinfo {author} {\bibfnamefont {S.-Y.}\ \bibnamefont
  {Li}}, \bibinfo {author} {\bibfnamefont {Y.-R.}\ \bibnamefont {Liu}},
  \bibinfo {author} {\bibfnamefont {Z.-L.}\ \bibnamefont {Man}}, \bibinfo
  {author} {\bibfnamefont {Z.-G.}\ \bibnamefont {Si}},\ and\ \bibinfo {author}
  {\bibfnamefont {J.}~\bibnamefont {Wu}},\ }\href
  {https://doi.org/10.1103/PhysRevD.108.056015} {\bibfield  {journal} {\bibinfo
   {journal} {Phys. Rev. D}\ }\textbf {\bibinfo {volume} {108}},\ \bibinfo
  {pages} {056015} (\bibinfo {year} {2023})},\ \Eprint
  {https://arxiv.org/abs/2307.00539} {arXiv:2307.00539 [hep-ph]} \BibitemShut
  {NoStop}%
\bibitem [{\citenamefont {Zhang}\ \emph {et~al.}(2024)\citenamefont {Zhang},
  \citenamefont {Liu},\ and\ \citenamefont {Jia}}]{Zhang:2023teh}%
  \BibitemOpen
  \bibfield  {author} {\bibinfo {author} {\bibfnamefont {W.-X.}\ \bibnamefont
  {Zhang}}, \bibinfo {author} {\bibfnamefont {C.-L.}\ \bibnamefont {Liu}},\
  and\ \bibinfo {author} {\bibfnamefont {D.}~\bibnamefont {Jia}},\ }\href
  {https://doi.org/10.1103/PhysRevD.109.114037} {\bibfield  {journal} {\bibinfo
   {journal} {Phys. Rev. D}\ }\textbf {\bibinfo {volume} {109}},\ \bibinfo
  {pages} {114037} (\bibinfo {year} {2024})},\ \Eprint
  {https://arxiv.org/abs/2312.12770} {arXiv:2312.12770 [hep-ph]} \BibitemShut
  {NoStop}%
\bibitem [{\citenamefont {Burns}\ and\ \citenamefont
  {Swanson}(2023)}]{Burns:2022uha}%
  \BibitemOpen
  \bibfield  {author} {\bibinfo {author} {\bibfnamefont {T.~J.}\ \bibnamefont
  {Burns}}\ and\ \bibinfo {author} {\bibfnamefont {E.~S.}\ \bibnamefont
  {Swanson}},\ }\href {https://doi.org/10.1016/j.physletb.2023.137715}
  {\bibfield  {journal} {\bibinfo  {journal} {Phys. Lett. B}\ }\textbf
  {\bibinfo {volume} {838}},\ \bibinfo {pages} {137715} (\bibinfo {year}
  {2023})},\ \Eprint {https://arxiv.org/abs/2208.05106} {arXiv:2208.05106
  [hep-ph]} \BibitemShut {NoStop}%
\bibitem [{\citenamefont {Clymton}\ \emph {et~al.}(2024)\citenamefont
  {Clymton}, \citenamefont {Kim},\ and\ \citenamefont
  {Mart}}]{Clymton:2024fbf}%
  \BibitemOpen
  \bibfield  {author} {\bibinfo {author} {\bibfnamefont {S.}~\bibnamefont
  {Clymton}}, \bibinfo {author} {\bibfnamefont {H.-C.}\ \bibnamefont {Kim}},\
  and\ \bibinfo {author} {\bibfnamefont {T.}~\bibnamefont {Mart}},\ }\href
  {https://doi.org/10.1103/PhysRevD.110.094014} {\bibfield  {journal} {\bibinfo
   {journal} {Phys. Rev. D}\ }\textbf {\bibinfo {volume} {110}},\ \bibinfo
  {pages} {094014} (\bibinfo {year} {2024})},\ \Eprint
  {https://arxiv.org/abs/2408.04166} {arXiv:2408.04166 [hep-ph]} \BibitemShut
  {NoStop}%
\bibitem [{\citenamefont {Ali}\ \emph {et~al.}(2019)\citenamefont {Ali} \emph
  {et~al.}}]{GlueX:2019mkq}%
  \BibitemOpen
  \bibfield  {author} {\bibinfo {author} {\bibfnamefont {A.}~\bibnamefont
  {Ali}} \emph {et~al.} (\bibinfo {collaboration} {GlueX}),\ }\href
  {https://doi.org/10.1103/PhysRevLett.123.072001} {\bibfield  {journal}
  {\bibinfo  {journal} {Phys. Rev. Lett.}\ }\textbf {\bibinfo {volume} {123}},\
  \bibinfo {pages} {072001} (\bibinfo {year} {2019})},\ \Eprint
  {https://arxiv.org/abs/1905.10811} {arXiv:1905.10811 [nucl-ex]} \BibitemShut
  {NoStop}%
\bibitem [{\citenamefont {Blankenbecler}\ and\ \citenamefont
  {Sugar}(1966)}]{Blankenbecler:1965gx}%
  \BibitemOpen
  \bibfield  {author} {\bibinfo {author} {\bibfnamefont {R.}~\bibnamefont
  {Blankenbecler}}\ and\ \bibinfo {author} {\bibfnamefont {R.}~\bibnamefont
  {Sugar}},\ }\href {https://doi.org/10.1103/PhysRev.142.1051} {\bibfield
  {journal} {\bibinfo  {journal} {Phys. Rev.}\ }\textbf {\bibinfo {volume}
  {142}},\ \bibinfo {pages} {1051} (\bibinfo {year} {1966})}\BibitemShut
  {NoStop}%
\bibitem [{\citenamefont {Aaron}\ \emph {et~al.}(1968)\citenamefont {Aaron},
  \citenamefont {Amado},\ and\ \citenamefont {Young}}]{Aaron:1968aoz}%
  \BibitemOpen
  \bibfield  {author} {\bibinfo {author} {\bibfnamefont {R.}~\bibnamefont
  {Aaron}}, \bibinfo {author} {\bibfnamefont {R.~D.}\ \bibnamefont {Amado}},\
  and\ \bibinfo {author} {\bibfnamefont {J.~E.}\ \bibnamefont {Young}},\ }\href
  {https://doi.org/10.1103/PhysRev.174.2022} {\bibfield  {journal} {\bibinfo
  {journal} {Phys. Rev.}\ }\textbf {\bibinfo {volume} {174}},\ \bibinfo {pages}
  {2022} (\bibinfo {year} {1968})}\BibitemShut {NoStop}%
\bibitem [{\citenamefont {Casalbuoni}\ \emph {et~al.}(1997)\citenamefont
  {Casalbuoni}, \citenamefont {Deandrea}, \citenamefont {Di~Bartolomeo},
  \citenamefont {Gatto}, \citenamefont {Feruglio},\ and\ \citenamefont
  {Nardulli}}]{Casalbuoni:1996pg}%
  \BibitemOpen
  \bibfield  {author} {\bibinfo {author} {\bibfnamefont {R.}~\bibnamefont
  {Casalbuoni}}, \bibinfo {author} {\bibfnamefont {A.}~\bibnamefont
  {Deandrea}}, \bibinfo {author} {\bibfnamefont {N.}~\bibnamefont
  {Di~Bartolomeo}}, \bibinfo {author} {\bibfnamefont {R.}~\bibnamefont
  {Gatto}}, \bibinfo {author} {\bibfnamefont {F.}~\bibnamefont {Feruglio}},\
  and\ \bibinfo {author} {\bibfnamefont {G.}~\bibnamefont {Nardulli}},\ }\href
  {https://doi.org/10.1016/S0370-1573(96)00027-0} {\bibfield  {journal}
  {\bibinfo  {journal} {Phys. Rept.}\ }\textbf {\bibinfo {volume} {281}},\
  \bibinfo {pages} {145} (\bibinfo {year} {1997})},\ \Eprint
  {https://arxiv.org/abs/hep-ph/9605342} {arXiv:hep-ph/9605342} \BibitemShut
  {NoStop}%
\bibitem [{\citenamefont {Isola}\ \emph {et~al.}(2003)\citenamefont {Isola},
  \citenamefont {Ladisa}, \citenamefont {Nardulli},\ and\ \citenamefont
  {Santorelli}}]{Isola:2003fh}%
  \BibitemOpen
  \bibfield  {author} {\bibinfo {author} {\bibfnamefont {C.}~\bibnamefont
  {Isola}}, \bibinfo {author} {\bibfnamefont {M.}~\bibnamefont {Ladisa}},
  \bibinfo {author} {\bibfnamefont {G.}~\bibnamefont {Nardulli}},\ and\
  \bibinfo {author} {\bibfnamefont {P.}~\bibnamefont {Santorelli}},\ }\href
  {https://doi.org/10.1103/PhysRevD.68.114001} {\bibfield  {journal} {\bibinfo
  {journal} {Phys. Rev. D}\ }\textbf {\bibinfo {volume} {68}},\ \bibinfo
  {pages} {114001} (\bibinfo {year} {2003})},\ \Eprint
  {https://arxiv.org/abs/hep-ph/0307367} {arXiv:hep-ph/0307367} \BibitemShut
  {NoStop}%
\bibitem [{\citenamefont {Kawarabayashi}\ and\ \citenamefont
  {Suzuki}(1966)}]{Kawarabayashi:1966kd}%
  \BibitemOpen
  \bibfield  {author} {\bibinfo {author} {\bibfnamefont {K.}~\bibnamefont
  {Kawarabayashi}}\ and\ \bibinfo {author} {\bibfnamefont {M.}~\bibnamefont
  {Suzuki}},\ }\href {https://doi.org/10.1103/PhysRevLett.16.255} {\bibfield
  {journal} {\bibinfo  {journal} {Phys. Rev. Lett.}\ }\textbf {\bibinfo
  {volume} {16}},\ \bibinfo {pages} {255} (\bibinfo {year} {1966})}\BibitemShut
  {NoStop}%
\bibitem [{\citenamefont {Riazuddin}\ and\ \citenamefont
  {Fayyazuddin}(1966)}]{Riazuddin:1966sw}%
  \BibitemOpen
  \bibfield  {author} {\bibinfo {author} {\bibnamefont {Riazuddin}}\ and\
  \bibinfo {author} {\bibnamefont {Fayyazuddin}},\ }\href
  {https://doi.org/10.1103/PhysRev.147.1071} {\bibfield  {journal} {\bibinfo
  {journal} {Phys. Rev.}\ }\textbf {\bibinfo {volume} {147}},\ \bibinfo {pages}
  {1071} (\bibinfo {year} {1966})}\BibitemShut {NoStop}%
\bibitem [{\citenamefont {Bardeen}\ \emph {et~al.}(2003)\citenamefont
  {Bardeen}, \citenamefont {Eichten},\ and\ \citenamefont
  {Hill}}]{Bardeen:2003kt}%
  \BibitemOpen
  \bibfield  {author} {\bibinfo {author} {\bibfnamefont {W.~A.}\ \bibnamefont
  {Bardeen}}, \bibinfo {author} {\bibfnamefont {E.~J.}\ \bibnamefont
  {Eichten}},\ and\ \bibinfo {author} {\bibfnamefont {C.~T.}\ \bibnamefont
  {Hill}},\ }\href {https://doi.org/10.1103/PhysRevD.68.054024} {\bibfield
  {journal} {\bibinfo  {journal} {Phys. Rev. D}\ }\textbf {\bibinfo {volume}
  {68}},\ \bibinfo {pages} {054024} (\bibinfo {year} {2003})},\ \Eprint
  {https://arxiv.org/abs/hep-ph/0305049} {arXiv:hep-ph/0305049} \BibitemShut
  {NoStop}%
\bibitem [{\citenamefont {Liu}\ and\ \citenamefont {Oka}(2012)}]{Liu:2011xc}%
  \BibitemOpen
  \bibfield  {author} {\bibinfo {author} {\bibfnamefont {Y.-R.}\ \bibnamefont
  {Liu}}\ and\ \bibinfo {author} {\bibfnamefont {M.}~\bibnamefont {Oka}},\
  }\href {https://doi.org/10.1103/PhysRevD.85.014015} {\bibfield  {journal}
  {\bibinfo  {journal} {Phys. Rev. D}\ }\textbf {\bibinfo {volume} {85}},\
  \bibinfo {pages} {014015} (\bibinfo {year} {2012})},\ \Eprint
  {https://arxiv.org/abs/1103.4624} {arXiv:1103.4624 [hep-ph]} \BibitemShut
  {NoStop}%
\bibitem [{\citenamefont {Yan}\ \emph {et~al.}(1992)\citenamefont {Yan},
  \citenamefont {Cheng}, \citenamefont {Cheung}, \citenamefont {Lin},
  \citenamefont {Lin},\ and\ \citenamefont {Yu}}]{Yan:1992gz}%
  \BibitemOpen
  \bibfield  {author} {\bibinfo {author} {\bibfnamefont {T.-M.}\ \bibnamefont
  {Yan}}, \bibinfo {author} {\bibfnamefont {H.-Y.}\ \bibnamefont {Cheng}},
  \bibinfo {author} {\bibfnamefont {C.-Y.}\ \bibnamefont {Cheung}}, \bibinfo
  {author} {\bibfnamefont {G.-L.}\ \bibnamefont {Lin}}, \bibinfo {author}
  {\bibfnamefont {Y.~C.}\ \bibnamefont {Lin}},\ and\ \bibinfo {author}
  {\bibfnamefont {H.-L.}\ \bibnamefont {Yu}},\ }\href
  {https://doi.org/10.1103/PhysRevD.46.1148} {\bibfield  {journal} {\bibinfo
  {journal} {Phys. Rev. D}\ }\textbf {\bibinfo {volume} {46}},\ \bibinfo
  {pages} {1148} (\bibinfo {year} {1992})},\ \bibinfo {note} {[Erratum:
  Phys.Rev.D 55, 5851 (1997)]}\BibitemShut {NoStop}%
\bibitem [{\citenamefont {Chen}\ \emph {et~al.}(2019)\citenamefont {Chen},
  \citenamefont {Sun}, \citenamefont {Liu},\ and\ \citenamefont
  {Zhu}}]{Chen:2019asm}%
  \BibitemOpen
  \bibfield  {author} {\bibinfo {author} {\bibfnamefont {R.}~\bibnamefont
  {Chen}}, \bibinfo {author} {\bibfnamefont {Z.-F.}\ \bibnamefont {Sun}},
  \bibinfo {author} {\bibfnamefont {X.}~\bibnamefont {Liu}},\ and\ \bibinfo
  {author} {\bibfnamefont {S.-L.}\ \bibnamefont {Zhu}},\ }\href
  {https://doi.org/10.1103/PhysRevD.100.011502} {\bibfield  {journal} {\bibinfo
   {journal} {Phys. Rev. D}\ }\textbf {\bibinfo {volume} {100}},\ \bibinfo
  {pages} {011502} (\bibinfo {year} {2019})},\ \Eprint
  {https://arxiv.org/abs/1903.11013} {arXiv:1903.11013 [hep-ph]} \BibitemShut
  {NoStop}%
\bibitem [{\citenamefont {Dong}\ \emph {et~al.}(2021)\citenamefont {Dong},
  \citenamefont {Guo},\ and\ \citenamefont {Zou}}]{Dong:2021juy}%
  \BibitemOpen
  \bibfield  {author} {\bibinfo {author} {\bibfnamefont {X.-K.}\ \bibnamefont
  {Dong}}, \bibinfo {author} {\bibfnamefont {F.-K.}\ \bibnamefont {Guo}},\ and\
  \bibinfo {author} {\bibfnamefont {B.-S.}\ \bibnamefont {Zou}},\ }\href
  {https://doi.org/10.13725/j.cnki.pip.2021.02.001} {\bibfield  {journal}
  {\bibinfo  {journal} {Progr. Phys.}\ }\textbf {\bibinfo {volume} {41}},\
  \bibinfo {pages} {65} (\bibinfo {year} {2021})},\ \Eprint
  {https://arxiv.org/abs/2101.01021} {arXiv:2101.01021 [hep-ph]} \BibitemShut
  {NoStop}%
\bibitem [{\citenamefont {Colangelo}\ \emph {et~al.}(2004)\citenamefont
  {Colangelo}, \citenamefont {De~Fazio},\ and\ \citenamefont
  {Pham}}]{Colangelo:2003sa}%
  \BibitemOpen
  \bibfield  {author} {\bibinfo {author} {\bibfnamefont {P.}~\bibnamefont
  {Colangelo}}, \bibinfo {author} {\bibfnamefont {F.}~\bibnamefont
  {De~Fazio}},\ and\ \bibinfo {author} {\bibfnamefont {T.~N.}\ \bibnamefont
  {Pham}},\ }\href {https://doi.org/10.1103/PhysRevD.69.054023} {\bibfield
  {journal} {\bibinfo  {journal} {Phys. Rev. D}\ }\textbf {\bibinfo {volume}
  {69}},\ \bibinfo {pages} {054023} (\bibinfo {year} {2004})},\ \Eprint
  {https://arxiv.org/abs/hep-ph/0310084} {arXiv:hep-ph/0310084} \BibitemShut
  {NoStop}%
\bibitem [{\citenamefont {Casalbuoni}\ \emph {et~al.}(1993)\citenamefont
  {Casalbuoni}, \citenamefont {Deandrea}, \citenamefont {Di~Bartolomeo},
  \citenamefont {Gatto}, \citenamefont {Feruglio},\ and\ \citenamefont
  {Nardulli}}]{Casalbuoni:1992fd}%
  \BibitemOpen
  \bibfield  {author} {\bibinfo {author} {\bibfnamefont {R.}~\bibnamefont
  {Casalbuoni}}, \bibinfo {author} {\bibfnamefont {A.}~\bibnamefont
  {Deandrea}}, \bibinfo {author} {\bibfnamefont {N.}~\bibnamefont
  {Di~Bartolomeo}}, \bibinfo {author} {\bibfnamefont {R.}~\bibnamefont
  {Gatto}}, \bibinfo {author} {\bibfnamefont {F.}~\bibnamefont {Feruglio}},\
  and\ \bibinfo {author} {\bibfnamefont {G.}~\bibnamefont {Nardulli}},\ }\href
  {https://doi.org/10.1016/0370-2693(93)91521-N} {\bibfield  {journal}
  {\bibinfo  {journal} {Phys. Lett. B}\ }\textbf {\bibinfo {volume} {309}},\
  \bibinfo {pages} {163} (\bibinfo {year} {1993})},\ \Eprint
  {https://arxiv.org/abs/hep-ph/9304280} {arXiv:hep-ph/9304280} \BibitemShut
  {NoStop}%
\bibitem [{\citenamefont {Shimizu}\ and\ \citenamefont
  {Harada}(2017)}]{Shimizu:2017xrg}%
  \BibitemOpen
  \bibfield  {author} {\bibinfo {author} {\bibfnamefont {Y.}~\bibnamefont
  {Shimizu}}\ and\ \bibinfo {author} {\bibfnamefont {M.}~\bibnamefont
  {Harada}},\ }\href {https://doi.org/10.1103/PhysRevD.96.094012} {\bibfield
  {journal} {\bibinfo  {journal} {Phys. Rev. D}\ }\textbf {\bibinfo {volume}
  {96}},\ \bibinfo {pages} {094012} (\bibinfo {year} {2017})},\ \Eprint
  {https://arxiv.org/abs/1708.04743} {arXiv:1708.04743 [hep-ph]} \BibitemShut
  {NoStop}%
\bibitem [{\citenamefont {Kim}\ \emph {et~al.}(1994)\citenamefont {Kim},
  \citenamefont {Durso},\ and\ \citenamefont {Holinde}}]{Kim:1994ce}%
  \BibitemOpen
  \bibfield  {author} {\bibinfo {author} {\bibfnamefont {H.-C.}\ \bibnamefont
  {Kim}}, \bibinfo {author} {\bibfnamefont {J.~W.}\ \bibnamefont {Durso}},\
  and\ \bibinfo {author} {\bibfnamefont {K.}~\bibnamefont {Holinde}},\ }\href
  {https://doi.org/10.1103/PhysRevC.49.2355} {\bibfield  {journal} {\bibinfo
  {journal} {Phys. Rev. C}\ }\textbf {\bibinfo {volume} {49}},\ \bibinfo
  {pages} {2355} (\bibinfo {year} {1994})}\BibitemShut {NoStop}%
\bibitem [{\citenamefont {Kim}\ and\ \citenamefont {Kim}(2018)}]{Kim:2018nqf}%
  \BibitemOpen
  \bibfield  {author} {\bibinfo {author} {\bibfnamefont {J.-Y.}\ \bibnamefont
  {Kim}}\ and\ \bibinfo {author} {\bibfnamefont {H.-C.}\ \bibnamefont {Kim}},\
  }\href {https://doi.org/10.1103/PhysRevD.97.114009} {\bibfield  {journal}
  {\bibinfo  {journal} {Phys. Rev. D}\ }\textbf {\bibinfo {volume} {97}},\
  \bibinfo {pages} {114009} (\bibinfo {year} {2018})},\ \Eprint
  {https://arxiv.org/abs/1803.04069} {arXiv:1803.04069 [hep-ph]} \BibitemShut
  {NoStop}%
\bibitem [{\citenamefont {Kim}\ \emph {et~al.}(2021)\citenamefont {Kim},
  \citenamefont {Kim}, \citenamefont {Yang},\ and\ \citenamefont
  {Oka}}]{Kim:2021xpp}%
  \BibitemOpen
  \bibfield  {author} {\bibinfo {author} {\bibfnamefont {J.-Y.}\ \bibnamefont
  {Kim}}, \bibinfo {author} {\bibfnamefont {H.-C.}\ \bibnamefont {Kim}},
  \bibinfo {author} {\bibfnamefont {G.-S.}\ \bibnamefont {Yang}},\ and\
  \bibinfo {author} {\bibfnamefont {M.}~\bibnamefont {Oka}},\ }\href
  {https://doi.org/10.1103/PhysRevD.103.074025} {\bibfield  {journal} {\bibinfo
   {journal} {Phys. Rev. D}\ }\textbf {\bibinfo {volume} {103}},\ \bibinfo
  {pages} {074025} (\bibinfo {year} {2021})},\ \Eprint
  {https://arxiv.org/abs/2101.10653} {arXiv:2101.10653 [hep-ph]} \BibitemShut
  {NoStop}%
\bibitem [{\citenamefont {Haftel}\ and\ \citenamefont
  {Tabakin}(1970)}]{Haftel:1970zz}%
  \BibitemOpen
  \bibfield  {author} {\bibinfo {author} {\bibfnamefont {M.~I.}\ \bibnamefont
  {Haftel}}\ and\ \bibinfo {author} {\bibfnamefont {F.}~\bibnamefont
  {Tabakin}},\ }\href {https://doi.org/10.1016/0375-9474(70)90047-3} {\bibfield
   {journal} {\bibinfo  {journal} {Nucl. Phys. A}\ }\textbf {\bibinfo {volume}
  {158}},\ \bibinfo {pages} {1} (\bibinfo {year} {1970})}\BibitemShut {NoStop}%
\bibitem [{\citenamefont {Machleidt}\ \emph {et~al.}(1987)\citenamefont
  {Machleidt}, \citenamefont {Holinde},\ and\ \citenamefont
  {Elster}}]{Machleidt:1987hj}%
  \BibitemOpen
  \bibfield  {author} {\bibinfo {author} {\bibfnamefont {R.}~\bibnamefont
  {Machleidt}}, \bibinfo {author} {\bibfnamefont {K.}~\bibnamefont {Holinde}},\
  and\ \bibinfo {author} {\bibfnamefont {C.}~\bibnamefont {Elster}},\ }\href
  {https://doi.org/10.1016/S0370-1573(87)80002-9} {\bibfield  {journal}
  {\bibinfo  {journal} {Phys. Rept.}\ }\textbf {\bibinfo {volume} {149}},\
  \bibinfo {pages} {1} (\bibinfo {year} {1987})}\BibitemShut {NoStop}%
\bibitem [{\citenamefont {Aaij}\ \emph {et~al.}(2021)\citenamefont {Aaij} \emph
  {et~al.}}]{LHCb:2020jpq}%
  \BibitemOpen
  \bibfield  {author} {\bibinfo {author} {\bibfnamefont {R.}~\bibnamefont
  {Aaij}} \emph {et~al.} (\bibinfo {collaboration} {LHCb}),\ }\href
  {https://doi.org/10.1016/j.scib.2021.02.030} {\bibfield  {journal} {\bibinfo
  {journal} {Sci. Bull.}\ }\textbf {\bibinfo {volume} {66}},\ \bibinfo {pages}
  {1278} (\bibinfo {year} {2021})},\ \Eprint {https://arxiv.org/abs/2012.10380}
  {arXiv:2012.10380 [hep-ex]} \BibitemShut {NoStop}%
\bibitem [{\citenamefont {Oller}\ and\ \citenamefont
  {Meissner}(2001)}]{Oller:2000fj}%
  \BibitemOpen
  \bibfield  {author} {\bibinfo {author} {\bibfnamefont {J.~A.}\ \bibnamefont
  {Oller}}\ and\ \bibinfo {author} {\bibfnamefont {U.~G.}\ \bibnamefont
  {Meissner}},\ }\href {https://doi.org/10.1016/S0370-2693(01)00078-8}
  {\bibfield  {journal} {\bibinfo  {journal} {Phys. Lett. B}\ }\textbf
  {\bibinfo {volume} {500}},\ \bibinfo {pages} {263} (\bibinfo {year}
  {2001})},\ \Eprint {https://arxiv.org/abs/hep-ph/0011146}
  {arXiv:hep-ph/0011146} \BibitemShut {NoStop}%
\bibitem [{\citenamefont {Jido}\ \emph {et~al.}(2003)\citenamefont {Jido},
  \citenamefont {Oller}, \citenamefont {Oset}, \citenamefont {Ramos},\ and\
  \citenamefont {Meissner}}]{Jido:2003cb}%
  \BibitemOpen
  \bibfield  {author} {\bibinfo {author} {\bibfnamefont {D.}~\bibnamefont
  {Jido}}, \bibinfo {author} {\bibfnamefont {J.~A.}\ \bibnamefont {Oller}},
  \bibinfo {author} {\bibfnamefont {E.}~\bibnamefont {Oset}}, \bibinfo {author}
  {\bibfnamefont {A.}~\bibnamefont {Ramos}},\ and\ \bibinfo {author}
  {\bibfnamefont {U.~G.}\ \bibnamefont {Meissner}},\ }\href
  {https://doi.org/10.1016/S0375-9474(03)01598-7} {\bibfield  {journal}
  {\bibinfo  {journal} {Nucl. Phys. A}\ }\textbf {\bibinfo {volume} {725}},\
  \bibinfo {pages} {181} (\bibinfo {year} {2003})},\ \Eprint
  {https://arxiv.org/abs/nucl-th/0303062} {arXiv:nucl-th/0303062} \BibitemShut
  {NoStop}%
\bibitem [{\citenamefont {Clymton}\ and\ \citenamefont
  {Kim}(2023)}]{Clymton:2023txd}%
  \BibitemOpen
  \bibfield  {author} {\bibinfo {author} {\bibfnamefont {S.}~\bibnamefont
  {Clymton}}\ and\ \bibinfo {author} {\bibfnamefont {H.-C.}\ \bibnamefont
  {Kim}},\ }\href {https://doi.org/10.1103/PhysRevD.108.074021} {\bibfield
  {journal} {\bibinfo  {journal} {Phys. Rev. D}\ }\textbf {\bibinfo {volume}
  {108}},\ \bibinfo {pages} {074021} (\bibinfo {year} {2023})},\ \Eprint
  {https://arxiv.org/abs/2305.14812} {arXiv:2305.14812 [hep-ph]} \BibitemShut
  {NoStop}%
\bibitem [{\citenamefont {Clymton}\ and\ \citenamefont
  {Kim}(2024)}]{Clymton:2024pql}%
  \BibitemOpen
  \bibfield  {author} {\bibinfo {author} {\bibfnamefont {S.}~\bibnamefont
  {Clymton}}\ and\ \bibinfo {author} {\bibfnamefont {H.-C.}\ \bibnamefont
  {Kim}},\ }\href {https://doi.org/10.1103/PhysRevD.110.114002} {\bibfield
  {journal} {\bibinfo  {journal} {Phys. Rev. D}\ }\textbf {\bibinfo {volume}
  {110}},\ \bibinfo {pages} {114002} (\bibinfo {year} {2024})},\ \Eprint
  {https://arxiv.org/abs/2409.02420} {arXiv:2409.02420 [hep-ph]} \BibitemShut
  {NoStop}%
\bibitem [{\citenamefont {Mei\ss{}ner}(2020)}]{Meissner:2020khl}%
  \BibitemOpen
  \bibfield  {author} {\bibinfo {author} {\bibfnamefont {U.-G.}\ \bibnamefont
  {Mei\ss{}ner}},\ }\href {https://doi.org/10.3390/sym12060981} {\bibfield
  {journal} {\bibinfo  {journal} {Symmetry}\ }\textbf {\bibinfo {volume}
  {12}},\ \bibinfo {pages} {981} (\bibinfo {year} {2020})},\ \Eprint
  {https://arxiv.org/abs/2005.06909} {arXiv:2005.06909 [hep-ph]} \BibitemShut
  {NoStop}%
\end{thebibliography}%
\bibliographystyle{apsrev4-2}

\end{document}